\documentclass[a4paper,fleqn,colorlinks,hyperfootnotes=true]{cas-dc}
\usepackage[numbers]{natbib}
\usepackage[figuresright]{rotating}
\usepackage{moreverb}
\usepackage{amsmath,amssymb,amsfonts}
\usepackage{algorithmic}
\usepackage{graphicx}
\usepackage{textcomp}
\usepackage{xcolor}
\usepackage{ragged2e}
\usepackage{balance}
\usepackage{colortbl}
\usepackage{multirow}
\usepackage{color}

\usepackage{CJKutf8}
\usepackage{siunitx}
\usepackage{rotfloat}
\usepackage{lscape}
\usepackage{amssymb,mathrsfs,amsmath}
\usepackage{subfigure}

\usepackage{pgfplots}
\usepackage{booktabs}
\usepackage{graphicx}

\usepackage{tcolorbox}
\usepackage{colortbl}
\usepackage{framed} % 引入framed宏包
\usepackage{listings}
\lstdefinestyle{mystyle}{
 backgroundcolor=\color{backcolor},
 commentstyle=\color{codegray},
 keywordstyle=\color{codeorange},
 numberstyle=\tiny\color{codegray},
 stringstyle=\color{codegreen},
 basicstyle=\ttfamily\footnotesize,
 breakatwhitespace=false,
 breaklines=true,
 captionpos=b,
 keepspaces=true,
 numbers=left,
 numbersep=5pt,
 showspaces=false,
 showstringspaces=false,
 showtabs=false,
 tabsize=2,
 xleftmargin=10pt,
}
\def\tsc#1{\csdef{#1}{\textsc{\lowercase{#1}}\xspace}}
\tsc{WGM}
\tsc{QE}
\tsc{EP}
\tsc{PMS}
\tsc{BEC}
\tsc{DE}
\begin{document}
\begin{sloppypar}

\let\WriteBookmarks\relax
\def\floatpagepagefraction{1}
\def\textpagefraction{.001}
\shorttitle{Code Clones in the Eclipse IIoT Software Ecosystem}
\shortauthors{Z Li et al.}
\title [mode = title]{Unveiling Code Clones in the Eclipse IIoT Software Ecosystem}

\author[1]{Zengyang Li}
\ead{zengyangli@ccnu.edu.cn}
\credit{Conceptualization, Methodology, Investigation, Data curation, Writing - Original draft preparation}

\address[1]{School of Computer Science \& Hubei Provincial Key Laboratory of Artificial Intelligence and Smart Learning, \\Central China Normal University, Wuhan, China \\}

\author[1]{Binbin Huang}
\ead{huangbinbin@mails.ccnu.edu.cn}
\credit{Conceptualization, Methodology, Investigation, Data curation, Software, Writing - Original draft preparation}

\author[1]{Yimeng Li}
\ead{liyimengccnu@mails.ccnu.edu.cn}
\credit{Methodology, Investigation, Data curation, Software}

\author[1]{Ran Mo}
\ead{moran@ccnu.edu.cn}
\credit{Conceptualization, Methodology, Writing - Original draft preparation}

\author[2]{Peng Liang}
\cormark[1]
\ead{liangp@whu.edu.cn}
\credit{Conceptualization, Methodology, Writing - Original draft preparation}
\address[2]{School of Computer Science, Wuhan University, Wuhan, China}

\author[3]{Hui Liu}
\ead{hliu@hust.edu.cn}
\credit{Methodology, Writing - Original draft preparation}
\address[3]{School of Artificial Intelligence and Automation, Huazhong University of Science and Technology, Wuhan, China}

\author[1]{Yutao Ma}
\ead{ytma@ccnu.edu.cn}
\credit{Conceptualization, Methodology, Writing - Original draft preparation}

\cortext[cor1]{Corresponding author.}

%\fundinginfo{Natural Science Foundation of Hubei
%Province of China, Grant Number: 2021CFB577\\
%National Natural Science Foundation of China, Grant Number: 62176099}

\begin{abstract}
\noindent\textbf{Context}: Industrial Internet of Things (IIoT) has become a prominent topic recently, with an increasing number of IIoT open-source software (OSS) projects emerging, also within the Eclipse Foundation. Code cloning is a common practice that can adversely affect software maintenance. In the IIoT OSS domain, developers frequently reuse code and configurations for efficiency, which can lead to code clone proliferation and maintenance challenges. However, the extent and effects of code clones in the IIoT OSS domain remain understudied.

\noindent\textbf{Objective}: This study aims to investigate the prevalence, evolution, and co-modification of code clones within the Eclipse IIoT OSS ecosystem.

\noindent\textbf{Methods}: We collected 90 release versions from 15 projects in the Eclipse IIoT OSS ecosystem, and investigated their code clone situations based on source code and change history using the NiCad tool and our custom analysis module. The investigation covered clone distribution, patterns, evolution trends, co-modified clones, and cross-project clones.

\noindent\textbf{Results}: 1) Code clones are prevalent in Eclipse IIoT OSS projects, with 16.3\% of code lines involved in clones - nearly twice the proportion observed in traditional OSS projects; 2) Most code clones occur between commits, while there are still a significant proportion of code clones that each clone pair happens within a commit; 3) Most Eclipse IIoT projects remain stable in clone numbers during version iterations; 4) An average of 0.17\% of the clones have been co-modified, which negatively affect maintenance; and 5) Cross-project clone pairs are prevalent, more in Java than in C projects, with rare co-modifications (0.02\%) only in Java projects.

\noindent\textbf{Conclusions}: We analyzed the distribution characteristics of code clones in Eclipse IIoT OSS projects, examining their specific situations, causes, and development trends. Our findings highlight the potential negative impacts of these clones on software maintenance, emphasizing the need to address these issues to improve overall software quality.
\end{abstract}

\begin{keywords}
Code Clone, Industrial Internet of Things, Open Source Software, Empirical Study
\end{keywords}

\maketitle

\section{Introduction}
\label{chap:intro}
%研究背景
The Industrial Internet of Things (IIoT) integrates Internet technology, edge computing \citep{Zhu2022GreenAF}, digital twin \citep{tao2022digital}, and artificial intelligence \citep{Tian2022ABM} to transform global industrial systems into intelligent networks \citep{Gabsi2024IntegratingAI}. By connecting equipment, factories, and supply chains, IIoT enhances the intelligence, optimization, and collaboration of production processes, improving overall manufacturing efficiency and competitiveness. However, as the sizes of the software systems supporting these IIoT systems continue to expand, challenges in software maintenance are becoming increasingly prominent \cite{shakya2022challenges}. Recent literature suggests that managing embedded software maintenance in the IIoT domain is a daunting task, requiring modern methodologies to ensure software quality and reliability \cite{Fariha2024ASystematicES}.

Code quality issues can substantially raise maintenance costs and may even lead to program crashes and data loss in high-reliability environments \cite{Alami2024FreeOS, Brstler2023DevelopersTA}. Among the various factors impacting software maintenance, code clones (i.e., identical or similar code blocks) are considered a major cause of project source code bloat \cite{Higo2018CodeClone}. However, code clones are not inherently detrimental to software quality \cite{Kapser2006CloneHarmful}. Existing studies show that they can serve legitimate purposes in software development \cite{Saini2018ClonedAN}, such as in safety-critical systems where cloning may be preferred over abstraction to ensure module isolation and minimize dependencies \cite{Kernel2012MutableProtection}. While cloning is a common practice to improve initial development efficiency, the change risk it introduces can make systems harder to maintain throughout their long-term evolution \cite{MONDAL2018BugProne}. For example, empirical studies have demonstrated that the maintenance effort for methods reusing Stack Overflow code blocks increases significantly when clones are present \cite{CHEN2024Reuse}.

%论文的研究动机
Studies have shown that code clone prevalence and patterns vary significantly across domains \cite{Mo2023ACS, Mo2025CodeCloneOnSSC, Zhao2022CCMicroservice}. Recent empirical studies further indicate that code clones are more likely to occur in deep learning software compared to traditional projects \cite{Mo2023ExploringTI}. Although code clones in general IoT systems have been investigated \cite{ullah2021clone, zhu2024multilingual}, these findings cannot be directly generalized to the IIoT context. Unlike consumer IoT, which primarily causes inconvenience, software defects propagated through clones in IIoT can lead to catastrophic safety hazards or significant economic losses \cite{Boyes2018TheII}. Furthermore, while general IoT research suggests a ``copy-and-forget'' pattern for cross-project clones where up to 95\% remain unmodified \cite{zhu2024multilingual}, it remains unclear whether this applies to the IIoT ecosystem, which relies on the long-term collaborative maintenance of standard industrial protocols.

Crucially, in safety-critical IIoT systems, stability often takes precedence over architectural abstraction \cite{IHIRWE2024IoT}. Developers are typically hesitant to modify working code solely for the purpose of reducing clones, as refactoring may introduce new risks and instability \cite{Pantiuchina2021Refactory}. Therefore, the objective of understanding code clones in this domain is not to force immediate refactoring, but rather to provide risk monitoring and consistency management. By detecting clones and quantifying co-modifications, developers can ensure that necessary updates are executed synchronously across duplicates, thereby reducing the ``ripple effect'' of changes without compromising system stability.

%采取的研究方法
Based on this, we focused our research on the Eclipse IIoT Open-Source Software (OSS) ecosystem, a long-standing and representative open-source IIoT collaborative community. Projects in this ecosystem implement de-facto industrial standards and possess a rich history of maintenance. According to the statistics of the collected projects in our dataset, the average project lifespan is approximately 9 years. This maturity provides a solid foundation for analyzing long-term evolutionary trends of code clones. Specifically, we aim to bridge the gap between general code clone research and the specific IIoT maintenance context. First, by quantifying clone prevalence and distribution, we assess the extent of code bloat and its subsequent overhead on constrained hardware resources as well as the extra maintenance effort required in IIoT environments. Second, by analyzing commit patterns and evolutionary trends, we reveal how code clones are introduced and managed over time, identifying potential stability risks during rapid iterations. Third, and most importantly regarding maintenance effort, we investigate co-modified clones to determine if they indeed necessitate simultaneous updates, thereby quantifying the ``ripple effect'' of changes. Finally, we examine cross-project clones to identify ecosystem-wide redundancy and propagation risks. Based on these empirical findings, we propose actionable implications to help developers optimize their code management strategies in IIoT software.

%以下为本文主要贡献
In this work, we performed an in-depth analysis of IIoT OSS projects in the Eclipse ecosystem. The main \textbf{contributions} of this work are summarized below:
\begin{itemize}
\setlength{\itemsep}{0pt}
\setlength{\parsep}{0pt}
    \item[$\bullet$] To our knowledge, this work is the first attempt to analyze code clones in projects from the Eclipse IIoT OSS ecosystem.
    \item[$\bullet$] We investigated the prevalence, distribution, and evolution of code clones to reveal the dynamics of code clone in IIoT projects.
    \item[$\bullet$] We analyzed the clone patterns (source and clone locations) to understand the way of clone introduction.
    \item[$\bullet$] We quantified the co-modification occurrences of the code clones to measure the actual maintenance effort required by duplicated code.
    \item[$\bullet$] We identified cross-project code clones and, for the first time, investigated their co-modification across projects to assess ecosystem-wide coupling risks.
\end{itemize}

%以下为论文结构
The remainder of this paper is organized as follows. Section \ref{sec_relatedwork} discusses the work related to our research, Section \ref{sec_RQ} describes our research design, Section \ref{sec_results} reports the results of our study, Section \ref{sec_discussion} discusses the study findings, Section \ref{sec_threats} presents the threats to the validity of the study results, and Section \ref{sec_conclusion} concludes this work with future research directions.

\section{Related Work}\label{sec_relatedwork}
We examined the related work in two aspects, i.e., IIoT software quality and code clone.

\subsection{IIoT Software Quality}
% 在 IIoT 领域，已经有不少学者对软件安全和软件质量进行了研究。Mugarza等人分析了 IEC 62443 工业安全标准，并提出IIoT系统和组件的安全维护管理是一项需要解决的艰巨任务；Landeck等人通过分析 IIoT 平台的架构和相应的利益相关者，建议研究人员开发将车间安全要求映射到信息技术保护措施的模型；并提出未来的工作应分析针对特定情境的软件安全评估的可行方法，以及 IIoT 安全的组成性；Sengupta等人探讨了集中式 IoT/IIoT 架构带来的挑战，并提出通过使用智能系统将物理世界与虚拟世界紧密耦合，这进一步加剧了基于IIoT的系统的漏洞；Fariha等人通过对不同需求、工程和维护活动所用方法的研究趋势的既定系统评价过程，对 79 项IIoT软件中有关嵌入式软件质量的主要研究进行了全面分析，并提出通过应用现代方法，可以达到缩短上市时间和提高质量的目的。
% 然而，上述研究大多是针对安全或性能的研究，没有专门针对IIoT软件维护的研究，因此我们选择IIoT软件维护为主要研究角度，考虑到代码克隆是影响软件维护的主要影响因素之一，所以我们决定在本文中从代码克隆的角度来研究IIoT软件维护。

In the IIoT domain, considerable research has been conducted on software quality (e.g., software security and performance). Mugarza \textit{et al.} analyzed the IEC 62443 industrial security standard and highlighted that the security maintenance management of IIoT systems and components is a daunting task that needs to be addressed \citep{Mugarza2020SecurityIA}. Landeck \textit{et al.} analyzed the architecture of IIoT platforms and the relevant stakeholders, suggesting that researchers develop models to map workshop safety requirements to information technology protection measures. They also proposed that future work should focus on analyzing feasible methods for software security assessment specific to certain contexts, as well as the composability of IIoT security \citep{Landeck2024SoftwareIT}. Sengupta \textit{et al.} explored the challenges brought by centralized IoT/IIoT architectures, pointing out that tightly coupling the physical and virtual worlds through intelligent systems further exacerbates vulnerabilities in IIoT-based systems \citep{Sengupta2020ACS}. Fariha \textit{et al.} conducted a comprehensive analysis of major studies related to embedded software quality across 79 IIoT software projects, using an established systematic review process to investigate trends in methods employed for different requirements, engineering, and maintenance activities. They proposed that by applying modern methodologies, it is possible to shorten time-to-market and improve quality \citep{Fariha2024ASystematicES}.

However, most of the aforementioned studies focus on security or performance, with no specific research dedicated to the maintainability of IIoT software. 
%Therefore, we chose IIoT software maintenance as our primary research perspective. 
In addition, code clones are one of the major factors affecting software maintenance \citep{Lozano2008AssessingTE, Mondal2012CCincreasemaintenance}. Therefore, in this study, we aimed to investigate the maintainability of IIoT software from the perspective of code clones.

\subsection{Code Clone}
% \textbf{Code Clone.}
\subsubsection{Code Clone Types and Terminology} \label{CCRelWork}
Code clone is the presence of duplicate or similar code blocks in the code \citep{juergens2009code}. This usually happens when programmers copy and paste code in different locations or reuse frameworks or design patterns \citep{Aversano2007HowCA}. In a way, code clones can increase developer efficiency and productivity by reusing existing function code or libraries; however, cloned code blocks may propagate changes and errors, which would make the software system harder to maintain~\citep{ALSHABIB2025SystemticCLSC, kim2005empirical, ZAKERINASRABADI2023SystematicSC}.

To facilitate the analysis of code clones, we introduce the following core concepts used throughout this study: 
\begin{itemize}
\setlength{\itemsep}{0pt}
\setlength{\parsep}{0pt}
 \item[$\bullet$] Code Block: A continuous segment of source code, specified by the file name, start line, and end line.
 \item[$\bullet$] Clone Pair: Two code blocks that are identical or similar according to the similarity criteria defined by the clone detection approach used in this study (see Section \ref{Code Clone Detection}). A clone pair is the basic unit of clone reporting. 
 \item[$\bullet$] Clone Class (or Group): A set of code blocks where every pair of blocks in the set forms a clone pair. 
\end{itemize}

Generally,  code clones are categorized into three primary types based on their similarity levels:
\begin{itemize}
\setlength{\itemsep}{0pt}
\setlength{\parsep}{0pt}
    \item[$\bullet$] Type-1 (Exact Clones): The two code blocks are identical clones except for the differences in spaces and comments.
    \item[$\bullet$] Type-2 (Renamed Clones): The two code blocks are identical or renamed clones except for variables, identifiers, types, text, layout, comments, blank spaces.
    \item[$\bullet$] Type-3 (Gapped Clones): Clones of two code blocks that are syntactically similar or have gaps, except for statement-level differences, adding or removing statements, using other identifiers, comments, layouts, types, texts, and spaces, etc.
\end{itemize}

\subsubsection{Code Clone Analysis}
The presence of code clones can lead to increased resource demands, which in turn raises maintenance overhead, as changes made to one code block may be propagated to all other clone instances of that code block. Therefore, many researchers are investigating the code clones present in source code and analyzing their static and dynamic characteristics to identify the potential threats they pose to software maintenance \citep{Lozano2008AssessingTE, Mo2023ExploringTI, Mondal2012CCincreasemaintenance}. For example, Lazano \textit{et al.} compared the maintenance effort of methods with clones with those without clones and suggested that certain methods appear to significantly increase their maintenance effort when clones are present \citep{Lozano2008AssessingTE}. Mo \textit{et al.} empirically investigated the impact of code clones on deep learning (DL) software in terms of prevalence and co-change and suggested that code clones are more likely to occur in DL software than in traditional projects and that a lot of co-changes occur in code clones in DL software \citep{Mo2023ExploringTI}.

However, to our knowledge, there has been no research on code clones in IIoT software so far, leaving the existence of code clones and their potential threats to maintainability in this type of software undiscovered. Therefore, in this study, we focus our analysis on code clones in IIoT software to verify whether code clones pose great maintenance challenges to IIoT software and the entire Eclipse IIoT OSS ecosystem.

\section{Study Design}\label{sec_RQ}
% RQ定义（做什么）
\subsection{Objective and Research Questions}
The goal of this study described using the Goal-Question-Metric (GQM) approach \citep{Ba1992} is: to analyze source code across various releases for the purpose of exploration with respect to the prevalence of code clones as well as their evolution and co-modification, from the point of view of software developers in the context of Eclipse IIoT OSS development.

Based on the above mentioned goal, we formulated five research questions (RQs). RQ1 and RQ2 investigate the state of code clones in terms of prevalence and commit distribution. RQ3 and RQ4 center around the evolution and co-modification characteristics of code clones and explore their possible impact on maintenance difficulty. RQ5 explores the existence and density of cross-project clones within the ecosystem.

%Code Clone相关的RQ
\noindent\textbf{RQ1: What is the presence of code clones in Eclipse IIoT OSS projects and how are they distributed?}
By detecting code clones and analyzing their distribution in each project, we can get a clear picture of the prevalence of code clones in different OSS projects in the Eclipse IIoT ecosystem. This can, in turn, raise developers' attention to aspects such as IIoT code refactoring and provide the developers with help on software maintenance.

%\noindent\textbf{RQ2: What are the clone patterns of code clones between different commits in the Eclipse IIoT OSS project?}
%\noindent\textbf{RQ2: What do the code clones distribute over the range of commits in the Eclipse IIoT OSS projects?}
\noindent\textbf{RQ2: How are code clones distributed across commits in Eclipse IIoT OSS projects?}
By examining whether the source and clone of each clone pair appear within the same commit or not, we can divide code clones into two code clone patterns, i.e., intra-commit clones and inter-commit clones. The distribution of the code clones over these two clone patterns can reveal the causes of code clones to some extent. This analysis enables differentiated governance: identifying intra-commit clones supports real-time intervention to intercept immediate duplication, while inter-commit clones distributions facilitate periodic monitoring to manage evolutionary code clones.

\noindent\textbf{RQ3: What are the evolutionary trends of the code clones in Eclipse IIoT OSS projects?}
Observing the number of code clones in recent versions, we can obtain the evolutionary trend of the code clones of the IIoT OSS projects. This allows for a more intuitive observation of changes in the number of code clones during project iterations, thereby revealing more clearly the history and future development direction of code clones in IIoT OSS projects. % This will improve the overall quality and development efficiency of IIoT OSS projects.

\noindent\textbf{RQ4: To what extent do code clones in Eclipse IIoT OSS projects involve co-modification?}
By observing and studying the percentage of co-modified code clones, it is possible to gain a deeper understanding of the impact of code clones on IIoT OSS projects, thus helping developers to better maintain their IIoT software.

\noindent\textbf{RQ5: Is there cross-project code clones in Eclipse IIoT OSS projects? If so, what is its distribution and density?}
Considering that the Eclipse IIoT software projects belong to the same OSS ecosystem, it is possible that these clones result from code reuse between the projects. This raises awareness among developers about the potential version synchronization issues and the risk of defect propagation that cross-project clones may cause.

% 项目挑选（在哪里做）
\subsection{Project Selection}
To construct the dataset for this study, we selected projects following the three criteria below:

\textbf{C1:} The project must belong to the Eclipse IIoT OSS ecosystem. The Eclipse Foundation hosts the largest community for open-source IIoT collaboration, managing over 50 projects with contributions from industry leaders like Bosch and Red Hat. These projects implement de-facto standards (e.g., MQTT, LwM2M). Therefore, we selected the Eclipse ecosystem as a representative sample of modern IIoT OSS development, rather than the entire proprietary IIoT landscape.

\textbf{C2:} The project must be currently active. To ensure the practical relevance of our maintenance recommendations, we focused on currently active projects (not archived and have had commits within a year). While we did not explicitly filter by project age during selection, our combination of activity and scale criteria implicitly selected mature projects with long-term evolution history, as verified in Table \ref{table:selectedEclipseIIoTOss}.

\textbf{C3:} The project must have more than 20K physical Lines of Code (LOC). The Eclipse IIoT ecosystem contains numerous small-scale ``example'' or ``demo'' repositories. To exclude these ``toy'' projects and ensure the selected subjects possess sufficient structural complexity to exhibit meaningful cloning patterns, we set a threshold of 20K LOC. This threshold serves as a heuristic to identify projects with substantial development effort and architectural maturity, distinguishing production-ready tools from simple tutorials. The LOC was measured using Cloc \citep{adanial_cloc} (excluding comments and blank lines).

Based on the above selection criteria, 60 OSS projects were initially obtained from the Eclipse IIoT ecosystem. After applying C2 and C3, 20 archived projects and 25 projects that did not meet the size threshold were removed. Finally, 15 projects (including 10 Java projects, 4 C projects, and 1 Python project) were included for data extraction and analysis. The basic information of these 15 selected projects is shown in Table \ref{table:selectedEclipseIIoTOss}.

\begin{table*}[h!]
    \centering
    \caption{Selected IIoT OSS Projects in Eclipse Ecosystem}
    \label{table:selectedEclipseIIoTOss}
    \scalebox{0.92}{
    \begin{tabular}{c c c c c c}
        \toprule
        \textbf{Project Name} & \textbf{Application Domain} & \textbf{Language} & \textbf{Start Date} & \textbf{Update Time} & \textbf{Versions}\\
        \midrule
        Arrowhead & Industrial Automation & Java & 2019-05-22 & 2024-03-19 & 4.4.0, 4.4.1, 4.5.0, 4.6.0, 4.6.1, 4.6.2 \\
        Californium & Cloud Service & Java & 2014-04-16 & 2024-08-21 & 3.8.0, 3.9.0, 3.9.1, 3.10.0, 3.11.0, 3.12.0 \\
        Cyclone DDS & IIoT Protocols & C & 2018-01-02 & 2024-08-22 & 0.9.1, 0.10.1, 0.10.2, 0.10.3, 0.10.4, 0.10.5 \\
        Ditto & Digital Twins & Java & 2017-04-10 & 2023-12-22 & 3.5.5, 3.5.6, 3.5.7, 3.5.8, 3.5.9, 3.5.10 \\
        Embedded CDT & Embedded IDE & C & 2013-09-21 & 2024-06-28 & 6.3.0, 6.3.1, 6.3.2, 6.4.0, 6.5.0, 6.6.0 \\
        Hawkbit & Edge Computing & Java & 2015-11-09 & 2024-08-23 & 0.3.0M7, 0.3.0M8, 0.3.0M9, 0.3.0, 0.4.1, 0.5.0 \\
        Hono & Cloud Service & Java & 2016-06-08 & 2024-04-20 & 2.2.0, 2.3.0, 2.4.0, 2.5.0, 2.5.1, 2.6.0 \\
        Kapua & Data Service & Java & 2016-10-14 & 2024-08-21 & 1.6.6, 1.6.7, 1.6.8, 1.6.9, 1.6.10, 1.6.11 \\
        Kura & Data Service & Java & 2013-12-16 & 2024-08-20 & 5.3.0, 5.3.1, 5.4.0, 5.4.1, 5.4.2, 5.5.0 \\
        Milo & IIoT Protocols & Java & 2016-05-06 & 2024-08-01 & 0.6.8, 0.6.9, 0.6.10, 0.6.11, 0.6.12, 0.6.13 \\
        Mosquitto & IIoT Protocols & C & 2014-05-06 & 2023-10-06 & 2.0.13, 2.0.14, 2.0.15, 2.0.16, 2.0.17, 2.0.18 \\
        Mraa & Edge Computing & C & 2014-04-08 & 2024-03-26 & 1.7.0, 1.8.0, 1.9.0, 2.0.0, 2.1.0, 2.2.0 \\
        Paho & IIoT Protocols & Java & 2012-03-26 & 2024-01-19 & 1.2.0, 1.2.1, 1.2.2, 1.2.3, 1.2.4, 1.2.5 \\
        VOLTTRON & Industrial Automation & Python & 2013-11-06 & 2024-05-08 & 8.1.1, 8.1.2, 8.1.3, 8.2.0, 9.0.0, 9.0.1 \\
        Vorto & Digital Twins & Java & 2015-03-02 & 2024-02-08 & 0.12.2, 0.12.3, 0.12.4, 0.12.5, 0.12.6, 1.0.0 \\
        \bottomrule
    \end{tabular}}
\end{table*}

% 数据采集（怎么做，得到什么）
\subsection{Data Collection}
% 收集什么（Data Items）
\subsubsection{Data Items to be Collected}\label{dataItemsCollected}
To answer the RQs, we take a clone pair as the unit of analysis and the data items to be collected were listed in Table \ref{table:dataItemsCollected}. All the data items to be collected except for D3 (ClonePattern), D5 (POC), and D9 (IsCoModified) are straightforward, thus we only explain these three data items in detail.

First, we explain data item D3 (ClonePattern). A clone pair consists of two code blocks, defined as the source block and the clone block. By extracting the commit logs and timestamps from the Git repository, we determine their relationship: 
\begin{itemize}
\setlength{\itemsep}{0pt}
\setlength{\parsep}{0pt}
    \item[$\bullet$] \textbf{Intra-commit clones (IaC)}: both blocks originate from the same commit.
    \item[$\bullet$] \textbf{Inter-commit clones (IeC)}: blocks come from different commits. 
\end{itemize}

Then, we give the definition of the Percentage of Cloned Code (POC) for a project (i.e., D5). Let $CLOC$ be the total physical lines of code involved in all identified clone blocks, where overlapping lines are counted only once. Let $SLOC$ be the total physical source lines of code of the project, excluding comments and blank lines. The POC is calculated as: 
\begin{equation}
    POC = \frac{CLOC}{SLOC} \times 100\%
\end{equation}

Finally, we explain data item D9 (IsCoModified). A clone pair is identified as co-modified if both code blocks of the pair are changed together across different versions. Suppose that a clone pair $C_1$ exists in version $V_1$. If the same pair is detected as $C_i$ in a subsequent version $V_i$, we compare the modified lines in both blocks. If both blocks in $C_i$ have undergone modifications to the same relative lines as $C_1$, we consider the clone to be co-modified.

\begin{table*}[]
    \caption{Data Items to be Collected for Each Clone Pair}
    \label{table:dataItemsCollected}
    \scalebox{1.0}{
    \begin{tabular}{cccc}
\hline
\textbf{\#} & \textbf{Name}  & \textbf{Description}                                            & \textbf{RQ}        \\ \hline
D1          & PairID         & The unique ID of the detected clone pair                        & RQ1-RQ5            \\
D2          & CloneType      & The similarity level (Type-1, Type-2, or Type-3)                & RQ1, RQ2, RQ4, RQ5 \\
D3          & ClonePattern   & Whether the blocks appear in the same commit (IaC) or not (IeC) & RQ2, RQ4           \\
D4          & CLOC           & The physical lines of code contained within clone blocks        & RQ1, RQ3           \\
D5          & POC            & Percentage of Cloned Code relative to the project size          & RQ1                \\
D6          & ProjectName    & The specific Eclipse IIoT project name                          & RQ1-RQ5            \\
D7          & Version        & The specific release version of the project                     & RQ3, RQ4           \\
D8          & CommitTime     & The timestamp of the commits in the Git history & RQ2                \\
D9          & IsCoModified   & Whether the clone pair was co-modified across versions          & RQ4                \\
D10         & IsCrossProject & Whether the clone pair exists between different projects        & RQ5                \\ \hline
\end{tabular}}
\end{table*}

% 用什么工具收（Tool）
\subsubsection{Code Clone Detection Tool Selection}\label{Code Clone Detection}
% 补充说明选择nicad的原因
% 在代码克隆检测领域，已开发出众多工具，涵盖从经典基于标记的检测器（如CCFinder \citep{Kamiya2002CCFinderAM}和Deckard \citep{Jiang2007DECKARDSA}）到可扩展技术（如SourcererCC \citep{Sajnani2015SourcererCCSC}）， 以及近期工具如NIL \citep{NIL2021}、CCSTokener \citep{CCSTokener2022}和MSCCD \citep{MSCCD2022}。为本实证研究选择最合适的工具，我们基于检测准确性（特别是对类型三克隆）、结果可解释性及对数据集规模的适应性对候选工具进行了评估。
% 我们确定NiCad \citep{Roy2008Nicad} 是我们特定研究情境下的最佳选择，主要基于以下三个原因：
% (1) 针对类型三克隆的高精度与鲁棒性：尽管新型工具主要侧重于超大规模仓库的可扩展性，NiCad仍保持着精度方面的黄金标准地位。在BigCloneBench数据集上的评估 \citep{Svajlenko2014TowardsAB} 表明，NiCad在所有克隆类型中可实现高达100%的召回率，并在适当配置下保持高精度（90%-96%）\citep{Sajnani2015SourcererCCSC, Roy2009FrameworkforEvaCC}。相比之下，CCFinder等工具对第三类克隆的召回率低于90%。尽管SourcererCC凭借93%的第三类召回率表现强劲，但NiCad基于TXL的解析技术能提供精细的结构分析，这对区分第三类克隆中微妙的语义差异至关重要。
% (2) 可解释性与可比性：不同于某些仅输出相似度评分的深度学习或"黑箱"方法，NiCad基于源代码规范化（美化打印）提供高度可解释的结果。此特性对本研究至关重要，因我们需通过详细人工检查分析共更改模式。此外，NiCad已在近期实证研究中广泛应用（如 \citep{Mo2023ExploringTI, Assi2024UnravelingCC}），确保我们的发现可与现有文献进行比较。采用成熟且经社区验证的工具，最大限度降低了实验工具偏见导致结果失真的风险。
% (3) 数据集规模适配性：我们注意到近期效率导向工具的进步，例如NIL \citep{NIL2021}和MSCCD \citep{MSCCD2025}专为高速处理海量多语言数据集而设计。但本研究数据集均为IIoT项目，所有代码行数均低于50万行。在此规模下，NiCad的执行时间完全可接受，使其无需像新型工具那样进行极端可扩展性权衡。相较于新型替代方案的原始速度，我们更重视NiCad对代码的精细准确解析能力。
% 因此，考虑到Eclipse IIoT开源项目的精度、可解释性和具体规模之间的平衡，我们选择了NiCad。如 \ref{CCRelWork} 节所述，我们的目标是检测 Type-1、Type-2 和 Type-3 克隆。NiCad能够通过特定参数配置识别这些类型。
% 我们为NiCad设置了三种配置，相关配置参数详见表\ref{table:configParaNiCad}。每种配置包含三个参数：用于区分T2与T1的标识符重命名参数、用于区分T2与T3的差异阈值参数（设定为30%），以及代码块的最小行数与最大行数参数（均设定为[5, 2500]）。最后一列标注了该配置可检测的代码克隆类型。这些参数在先前研究中已展现出检测精度与召回率的最优平衡点 \citep{Jebnoun2021ClonesID, Svajlenko2014CloneDetect}。

In the field of code clone detection, numerous detection tools have been developed and widely employed, ranging from classic token-based detectors like CCFinder \citep{Kamiya2002CCFinderAM} and Deckard \citep{Jiang2007DECKARDSA} to scalable techniques like SourcererCC \citep{Sajnani2015SourcererCCSC}, and more recent tools such as NIL \citep{NIL2021}, CCSTokener \citep{CCSTokener2023}, and MSCCD \citep{MSCCD2025}. To select the most appropriate tool for our empirical study, we evaluated candidates based on detection accuracy (especially for Type-3 clones), result interpretability, and execution efficiency as summarized in Table \ref{table:compareTools}.

We identified NiCad \citep{Roy2008Nicad} as the optimal choice for our specific research context for three primary reasons:

\textbf{(1) High Precision and Recall for Type-3 Clones}: While newer tools focus heavily on scalability for ultra-large repositories, NiCad remains a gold standard for precision. Evaluations on the BigCloneBench dataset \citep{Svajlenko2014TowardsAB} demonstrate that NiCad achieves up to 100\% recall across all clone types and maintains high precision (90\%-96\%) with appropriate configurations \citep{Roy2009FrameworkforEvaCC, Sajnani2015SourcererCCSC}. In contrast, tools like CCFinder drop below 90\% recall for Type-3 clones. Although SourcererCC is a strong competitor with high recall (93\% for Type-3), NiCad's TXL-based parsing offers fine-grained structural analysis that is crucial for distinguishing subtle semantic variations in Type-3 clones.

\textbf{(2) Result Interpretability}: Unlike some recent deep-learning-based or ``black-box'' approaches that output only similarity scores, NiCad provides highly interpretable results based on source code normalization (pretty-printing). This feature is essential for our study, as we require detailed manual inspection to analyze co-modification patterns. %Furthermore, NiCad has been extensively used in recent empirical studies (e.g., \citep{Assi2024UnravelingCC, Mo2023ExploringTI}), ensuring that our findings are comparable with existing literature. Using a mature, community-validated tool minimizes the risk that our observations are artifacts of experimental tool bias.

\textbf{(3) Execution Efficiency}: We acknowledged recent advancements in efficiency-oriented tools. For instance, NIL \citep{NIL2021} and MSCCD \citep{MSCCD2025} are designed to handle massive, multi-language datasets at speed. However, our dataset consists of IIoT projects that are all under 500K LOC (Lines of Code). At this scale, NiCad's execution time is entirely acceptable, making the extreme scalability trade-offs offered by newer tools unnecessary for our context. We prioritize the detailed, accurate parsing of NiCad over the raw speed of newer alternatives.

Consequently, considering the balance between precision, interpretability, and the execution efficiency, we selected NiCad. An additional benefit is that NiCad has been extensively used in recent empirical studies (e.g., \cite{Assi2024UnravelingCC, Mo2023ExploringTI}), ensuring that our findings are comparable with existing literature. As described in Section \ref{CCRelWork}, we aim to detect Type-1, Type-2, and Type-3 clones. NiCad is capable of identifying these types through specific parameter configurations.

\begin{table*}[h!]
    \centering
    \caption{Comparison of Different Code Clone Detection Tools}
    \label{table:compareTools}
    \scalebox{1.0}{
    \begin{tabular}{cccccc}
    \hline
    \multirow{2}{*}{\textbf{Tools}} & \multirow{2}{*}{\textbf{Analysis Method}} & \multirow{2}{*}{\textbf{Interpretability}} & \multirow{2}{*}{\textbf{Efficiency}} & \multicolumn{2}{c}{\textbf{Accuracy (Type-3)}} \\ \cline{5-6}
     &  &  &  & \textbf{Precision} & \textbf{Recall} \\ \hline
    \textbf{NiCad} & TXL-based & High (Pretty-print) & Medium & High (90\%-96\%) & $\sim$100\% \\
    \textbf{SourcererCC} & Token-based & Medium (Score) & High & High (91\%) & 93\% \\
    \textbf{NIL/MSCCD} & DL / Embedding & Low (Black-box) & Ultra-high & Variable & Variable \\
    \textbf{CCFinder} & Token-based & Medium & High & Moderate & \textless{}90\% \\ \hline
    \end{tabular}}
\end{table*}

We set up three configurations for NiCad to detect Type-1 (T1), Type-2 (T2), and Type-3 (T3) code clones, and our configuration parameters are shown in Table \ref{table:configParaNiCad}. Each configuration has three parameters, namely Identifier Renaming to distinguish T2 from T1, Dissimilarity Threshold to distinguish T2 from T3, which we set it at 30\%, and minimum lines of code and maximum lines of code for the code block, all of which we set to [5, 2500], and the last column is the type of code clone that this configuration can detect. These parameters have shown maximized detection precision and recall in previous studies \citep{Jebnoun2021ClonesID, Svajlenko2014CloneDetect}.

\begin{table}[h!]
    \centering
    \caption{The Configuration Settings of NiCad}
    \label{table:configParaNiCad}
    \scalebox{1.0}{
    \begin{tabular}{c c c c c}
         \toprule
         & Renaming & Threshold & LOC & Meaning \\
         \midrule
         config1 & none & 0\% & [5, 2500] & Type-1 (T1) \\
         config2 & blind & 0\% & [5, 2500] & Type-2 (T2) \\
         config3 & blind & 30\% & [5, 2500] & Type-3 (T3) \\
         \bottomrule
    \end{tabular}}
\end{table}

\subsubsection{Co-modified Code Clone Detection}
A so-called co-modified code clone is that: when a code block is changed, the clone of the block is also changed, as described in Section \ref{dataItemsCollected}, and this change increases the overhead of code maintenance \citep{Mo2023ACS, Mo2023ExploringTI}. To detect this particular class of clones, we developed an updated version (i.e., CCDetector-Update) based on CCDetector that has played a good role in detecting co-modified clones in deep learning software \citep{Mo2023ExploringTI}. In each project, we selected five previous versions as comparisons to detect co-modified clones. We then deduplicated the results to remove any duplicate clone pairs that may have appeared across the comparisons. Finally, we categorized the remaining results into three distinct types of clones.

In the process of selecting versions, we focused on release-level granularity for the evolutionary and co-modification analysis. While commits capture micro-changes, they often contain development ``noise'' that do not reflect the stable state of the software delivered to end-users. Such as temporary fixes, partial implementations, or reverted changes. Release versions represent verified milestones, reflecting the persistent technical debt and maintenance outcomes that project maintainers must support long-term. Focusing on releases allows us to observe the macro-trends of clone evolution and co-modification without being obscured by transient development fluctuations.

We selected six adjacent release versions for each project to form a continuous evolutionary timeline. To ensure ambiguity-free selection, we defined the ``latest release version'' as the most recent official release available as of our data collection cutoff date on August 30, 2024. Based on this anchor point, we traced back to the five immediately preceding releases. This window allows us to observe the continuity of code evolution across verified stable points. The specific versions selected for each project are detailed in the \textit{Versions} column of Table \ref{table:selectedEclipseIIoTOss}.

% 具体怎么收（Procedure）
\subsubsection{Data Collection Procedure}\label{Study Procedures}
\begin{figure*}[h]
    \centering
    \includegraphics[width=1\textwidth]{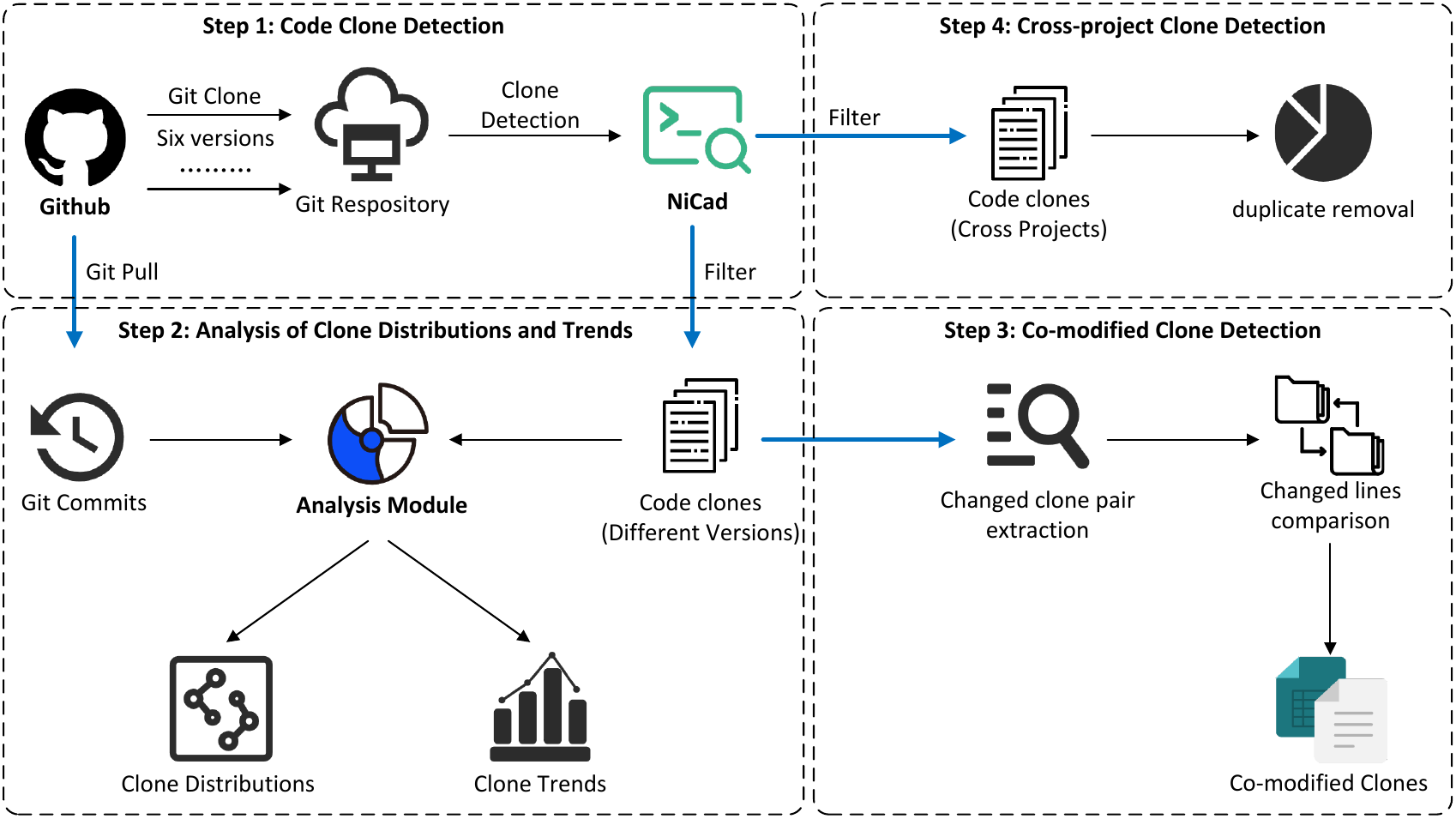}
    %\caption{Steps of our study}
    \caption{Overview of the research workflow, illustrating the main steps of the study, including dataset construction, clone detection, and subsequent analysis}
    \label{figure:expSteps}
\end{figure*}

We conducted our study by following the four main steps shown in Figure \ref{figure:expSteps}:

Step 1: \textit{Code Clone Detection}. We cloned the source code of the 15 projects from their GitHub repositories using $git clone$, obtaining the latest release version as the base. We then recursively fetching the five preceding adjacent release versions. We then employed NiCad \citep{Roy2008Nicad} to detect code clones for the source code of all 90 releases in total. Each clone was categorized using the three configurations described in Section \ref{Code Clone Detection} to obtain three types of clone pairs, while filtering out irrelevant information related to test and sample code. Since higher types of clones may obscure lower types (due to their lower similarity), we deduplicated all three types of clone pairs to derive a total count, which was then used for Step 2.

Step 2: \textit{Clone Distributions and Trends Analysis}. For the data obtained in Step 1, we combined the commit records of the same project with its six versions of clone pairs. We designed and implemented an analysis module to perform data analysis, which allowed us to examine the clone situation between different commits within the same project. Using various commit timestamps and records, we categorized clones into two patterns: those whose source and clone appear within a single commit (referred to as intra-commit clones) and those whose source and clone occur in different commits (referred to as inter-commit clones). In addition, we analyzed the evolutionary trends of the clone pairs within the same project as the iterated versions.

\begin{figure*}[h]
\centering
    \subfigure[Hono-2.5.1-ActilityEnterpriseProvider.java]
    {
        \includegraphics[width=0.48\textwidth]{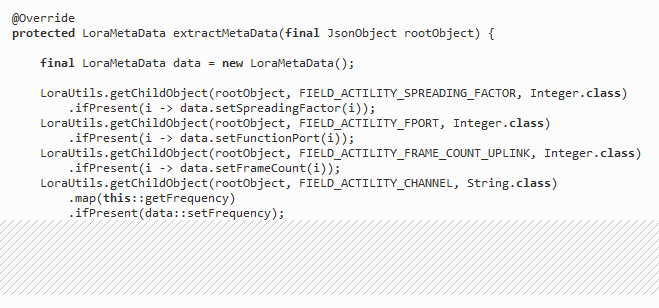}
    }
    \subfigure[Hono-2.6.0-ActilityEnterpriseProvider.java]
    {
        \includegraphics[width=0.48\textwidth]{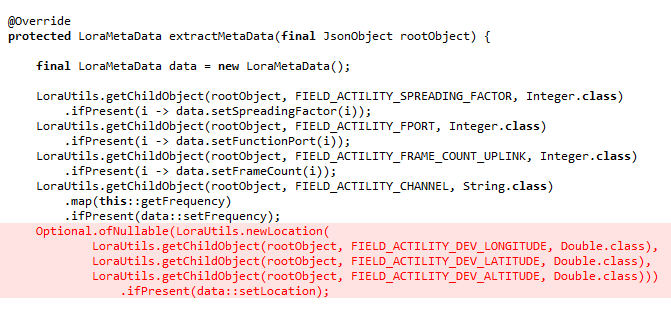}
    }
    \subfigure[Hono-2.5.1-ActilityWirelessProvider.java]
    {
        \includegraphics[width=0.48\textwidth]{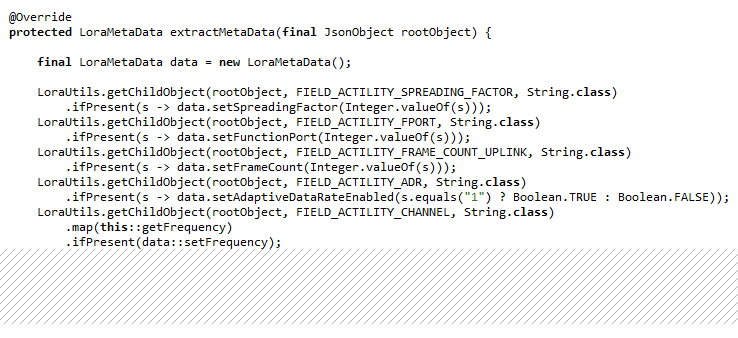}
    }
    \subfigure[Hono-2.6.0-ActilityWirelessProvider.java]
    {
        \includegraphics[width=0.48\textwidth]{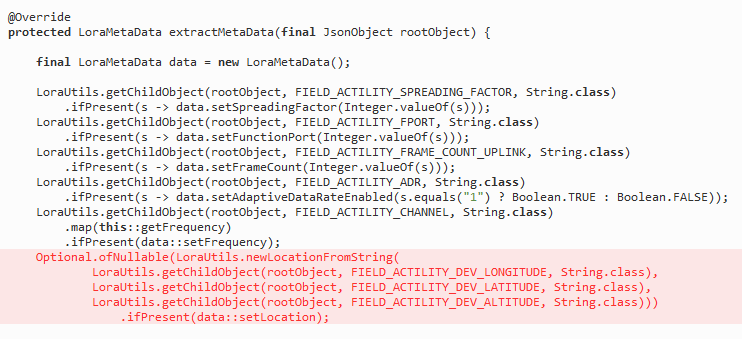}
    }
    \caption{Example co-modified code clones ((a) and (c) form a clone pair, while (b) and (d) form another clone pair).}
    \label{figure:exampleCoModifiedCodeClone}
\end{figure*}

Step 3: \textit{Co-modified Clone Detection}. We extracted the clone detection results of six versions of a project and compare them with the base version, using the last version as the base version. We compared clone results across adjacent versions. If both blocks in a pair underwent synchronized changes as described in Section \ref{dataItemsCollected}, they were marked as co-modified, as shown in Figure \ref{figure:exampleCoModifiedCodeClone}. In this step, all co-modified clones are gathered, and duplicate co-modified clone pairs are removed.

Step 4: \textit{Cross-project Clone Detection}. After obtaining clone situations among all projects, we combined the data from all projects and conducted another round of code clone detection using NiCad. We filtered out the clone pairs that belonged to the same project, retaining only the clone pairs between different projects. Subsequently, we removed irrelevant data related to tests and examples, resulting in a comprehensive overview of cross-project code clone situations.

% 实验步骤图

% 共更改克隆示例代码图

% 怎么分析
\subsection{Data Analysis} \label{sec:dataAnalysis} 
The answers to RQ1 and RQ2 can be obtained by descriptive statistics, focusing on the frequency and distribution of code clones within projects and across commits.

To answer RQ3, we analyzed the evolutionary trends of code clones across six release versions. We applied the Spearman's rank correlation test ($\rho$) to check the strength and direction of the monotonic relationship between the chronological version order and the number of clone pairs. In our case, the chronological version order ($X\in[1,6]$) and the number of clone pairs ($Y$). We interpreted the results as follows:
\begin{itemize}
\setlength{\itemsep}{0pt}
\setlength{\parsep}{0pt}
    \item[$\bullet$] A positive coefficient ($\rho\textgreater{0}$) with statistical significance ($p\textless{0.05}$) indicates a reliable Increasing trend (clones accumulate over time).
    \item[$\bullet$] A negative coefficient ($\rho\textless{0}$) with statistical significance ($p\textless{0.05}$) indicates a reliable Decreasing trend (clones are actively removed).
    \item[$\bullet$] A result with no statistical significance ($p\geq0.05$) indicates a stable equilibrium or fluctuating trend, suggesting that the introduction and removal of clones have reached a dynamic equilibrium. 
\end{itemize}

To answer RQ4, the occurrences of co-modified clones were analyzed to observe the impact of code clones on maintenance effort. We compared the co-modification rates between different clone types (T1, T2, and T3) and patterns (IaC and IeC) to identify which categories are more prone to synchronized changes.

To answer RQ5, we combined the data from all 15 projects to detect cross-project code clones. We identified the top 10 project pairs with the highest clone density for different clone types and investigated the likelihood of cross-project co-modifications by analyzing the last six release versions of all projects. This allows us to evaluate the risk of ecosystem-wide coupling and defect propagation.

\section{Study Results}\label{sec_results}
\subsection{Presence and Distribution of Code Clones in Eclipse IIoT OSS Projects (RQ1)}
To answer RQ1, we first employed NiCad to identify code clones from the latest release version of each project in a dataset of 15 IIoT OSS projects. Subsequently, we conducted a series of quantitative analysis based on the metrics defined in Section \ref{dataItemsCollected}, and the analysis results are summarized in Table \ref{table:cloneAnalysisEclipseIIoT} and Table \ref{table:PercentOfAll}.

\begin{table*}[h!]
    \centering
    \caption{Results of Clone Analysis for Eclipse IIoT OSS Projects (RQ1)}
    \label{table:cloneAnalysisEclipseIIoT}
    \scalebox{1.0}{
    \begin{tabular}{c|cccc|cc|cc|cc}
\hline
\multirow{2}{*}{\textbf{Project}} & \multicolumn{4}{c|}{\textbf{Total}}           & \multicolumn{2}{c|}{\textbf{Type-1}} & \multicolumn{2}{c|}{\textbf{Type-2}} & \multicolumn{2}{c}{\textbf{Type-3}} \\ \cline{2-11} 
                                  & \textbf{Pairs} & \textbf{SLOC} & \textbf{CLOC} & \textbf{POC} & \textbf{Pairs}    & \textbf{CLOC1}   & \textbf{Pairs}  & \textbf{CLOC2}   & \textbf{Pairs}   & \textbf{CLOC3}   \\ \hline
Arrowhead                         & 27,668 & 168,225         & 30,079        & 17.88\%      & 19,583            & 2,579            & 1,979             & 9,137            & 6,106            & 18,363           \\
Californium                       & 1,116 & 140,957          & 11,372        & 8.07\%       & 39                & 631              & 237               & 2,552            & 840              & 8,189            \\
Cyclone DDS                       & 1,600 & 225,334          & 16,612        & 7.37\%       & 15                & 439              & 389               & 3,754            & 1,196            & 12,419           \\
Ditto                             & 175,049 & 417,920        & 56,675        & 13.56\%      & 157               & 1,472            & 47,304            & 23,369           & 127,588          & 31,834           \\
Embedded CDT                      & 4,804 & 331,968          & 14,514        & 4.37\%       & 4,320             & 7,255            & 76                & 1,336            & 408              & 5,923            \\
Hawkbit                           & 1,658 & 90,008          & 7,205         & 8.00\%       & 23                & 281              & 723               & 2,581            & 912              & 4,343            \\
Hono                              & 1,076 & 118,836          & 7,912         & 6.66\%       & 20                & 259              & 309               & 2,220            & 746              & 5,433            \\
Kapua                             & 20,240 & 218,955         & 65,254        & 29.80\%      & 480               & 7,787            & 9,197             & 18,000           & 10,563           & 39,467           \\
Kura                              & 12,428 & 318,605         & 71,972        & 22.59\%      & 679               & 8,595            & 3,694             & 22,067           & 8,055            & 41,310           \\
Milo                              & 1,085,504 & 170,967      & 86,597        & 50.65\%      & 322               & 12,624           & 690,159           & 34,834           & 395,023          & 39,139           \\
Mosquitto                         & 420 & 46,197            & 7,326         & 15.86\%      & 16                & 185              & 100               & 1,562            & 304              & 5,579            \\
Mraa                              & 382 & 27,869            & 7,154         & 25.67\%      & 2                 & 54               & 114               & 1,579            & 266              & 5,521            \\
Paho                              & 784 & 34,470            & 15,226        & 44.17\%      & 199               & 4,672            & 180               & 2,375            & 405              & 8,179            \\
VOLTTRON                          & 145 & 131,339            & 2,296         & 1.75\%       & 16                & 105              & 44                & 620              & 85               & 1,571            \\
Vorto                             & 1,008 & 53,177          & 6,362         & 11.96\%      & 82                & 1,537            & 186               & 1,479            & 740              & 3,346            \\ \hline
\end{tabular}
    }
\end{table*}

\begin{table*}[h!]
 \centering
    \caption{Percentage of overall code clones in lines of code (RQ1)}
    \label{table:PercentOfAll}
    \scalebox{1.0}{
    \begin{tabular}{c c c c c c c c c}
    \hline
    \textbf{SLOC} & \textbf{CLOC} & \textbf{POC (All)} & \textbf{CLOC1} & \textbf{POC} & \textbf{CLOC2} & \textbf{POC} & \textbf{CLOC3} & \textbf{POC} \\ \hline
    2,494,827 & 406,556 & 16.30\% & 48,475 & 11.92\% & 127,465 & 31.35\% & 230,616 & 56.72\% \\ \hline
    \end{tabular}
    }
\end{table*}

Take project \texttt{Arrowhead} in Table \ref{table:cloneAnalysisEclipseIIoT} as an example. The results of \texttt{Arrowhead} show that: 1) There are 27,668 pairs of code clone (pairs) identified, with all clone blocks comprising a total of 30,079 $CLOC$, representing 17.88\% $POC$ within the project. 2) Regarding the detailed information, we note that 19,583, 1,979, and 6,106 clone pairs were identified for the T1, T2, and T3 clone, respectively, encompassing 2,579, 9,137, and 18,363 lines of code. These results suggest that code clones are prevalent in this project, with T3 being the most likely type of code clones to occur.

For all projects, we can summarize the following: 1) Code clones have been found in all projects. Project \texttt{Milo} has the highest number of 1,085,504 clone pairs, while project \texttt{VOLTTRON} has the least of 145 pairs. On average, 16.30\% (i.e., the proportion of the total cloned lines of code to the total lines of code) of source code lines being cloned. 2) Among all the cloned code, the three types of clones accounted for 11.92\%, 31.35\%, and 56.72\% of the total cloned lines, respectively. As shown in Table \ref{table:PercentOfAll}. These results indicate that code clones are prevalent in the Eclipse IIoT OSS projects, in which T3 clones exhibit a high incidence rate.

% 补充说明代码行问题
% 需要特别注意的是，克隆对的数量可能大于克隆代码行数（CLOC）。这是因为单个代码块可能同时参与多个克隆对，与不同代码块形成配对。因此，相同的代码行可能在不同配对中被多次计数，导致克隆对总数超过克隆代码行数。
% 特别是在Ditto和Milo项目中，克隆对数量（pairs）远大于克隆代码行数（CLOC）。这是因为这两个项目都包含大量高度相似的函数，从而形成了大型克隆簇——即每个代码块都与其他块相似的代码块集合。在这样的簇中，每个函数都会与所有其他相似函数形成克隆对，导致克隆对数量呈组合式增长。因此，尽管克隆行总数（CLOC）有限，但由于这些相似代码块之间存在密集的相互关联，克隆对数量仍会快速增加。
It is important to note that the number of clone pairs can be greater than the number of $CLOC$. This is because a single code block may participate in multiple clone pairs simultaneously, forming pairs with different code blocks. As a result, when there are many similar code blocks, the number of clone pairs formed among them can exceed the total number of $CLOC$, since every pairwise combination constitutes a separate clone pair.

In particular, for the \texttt{Ditto} and \texttt{Milo} projects, the number of clone pairs is much greater than the number of $CLOC$. This is because both projects contain a large number of highly similar functions, resulting in the formation of large clone clusters — groups of code blocks where each block is similar to others. Within such a cluster, each function forms clone pairs with all other similar functions, leading to a combinatorial increase in the number of clone pairs. As a result, even though the total number of $CLOC$ is limited, the number of clone pairs increases rapidly due to the dense interconnections among these similar code blocks.

\begin{tcolorbox}
\textbf{Finding 1}: Code clones are prevalent in Eclipse IIoT OSS projects. For an individual project, the proportion of cloned code ranges from 1.75\% to 50.65\%, while for all projects as a whole, the proportion of cloned code is 16.30\%. Type-3 clones make up 56.72\% of total cloned lines. Notably, over half of the lines of code in project \texttt{Milo} contain clones, with a total of 1.08 million clone pairs.
\end{tcolorbox}

\subsection{Code clone patterns (RQ2)}
To further examine the distribution of these clone pairs, we categorized them according to commit timing as described in Step 2 of Section \ref{Study Procedures}, identifying two distinct code clone patterns as described in \ref{dataItemsCollected}. We extracted the timestamp of each source and clone blocks from the commit log, thus determining each pair’s clone pattern. Table \ref{table:clonePairsDistribution} and Table \ref{table:PercentClonePairsDistribution} report the counts and percentages of Intra-commit clones and Inter-commit clones across the three clone types (T1–T3).

By examining Table \ref{table:clonePairsDistribution} and Table \ref{table:PercentClonePairsDistribution}, we observed that inter-commit clones dominate in most projects, with an average proportion of 56.97\%-76.78\% among all detected clone types. Nevertheless, intra-commit clones still represent a nontrivial share, with some projects showing particularly high proportions, for instance, intra-commit T1 clones reach as high as 74.53\% in project \texttt{Milo}.

Interestingly, the proportions of intra- and inter-commit clones vary significantly across projects. For example, in project \texttt{Milo}, intra-commit T1 clones are particularly frequent, indicating intensive copy-paste practices during development. Conversely, in project \texttt{Arrowhead}, inter-commit clones are more common. This difference could be potentially attributed to factors such as project size, modularity, team collaboration patterns, or the project's lifecycle stage, all of which can influence code reuse and maintenance activities across multiple commits and modules.

\begin{table*}[h!]
    \centering
    \caption{Distribution of Clone Pairs over Clone Patterns (RQ2)}
    \label{table:clonePairsDistribution}
    \scalebox{0.95}{
    \begin{tabular}{c|cc|ccc|ccc|ccc}
    \hline
    \multirow{2}{*}{\textbf{Project}} & \multicolumn{2}{c|}{\textbf{Total}} & \multicolumn{3}{c|}{\textbf{Type-1}} & \multicolumn{3}{c|}{\textbf{Type-2}} & \multicolumn{3}{c}{\textbf{Type-3}} \\ \cline{2-12} 
     & \textbf{IaC} & \textbf{IeC} & \textbf{All} & \textbf{IaC} & \textbf{IeC} & \textbf{All} & \textbf{IaC} & \textbf{IeC} & \textbf{All} & \textbf{IaC} & \textbf{IeC} \\ \hline
    Arrowhead & 3,074 & 24,594 & 19,583 & 1,877 & 17,706 & 1,979 & 491 & 1,488 & 6,106 & 706 & 5,400 \\
    Californium & 408 & 708 & 39 & 10 & 29 & 237 & 132 & 105 & 840 & 266 & 574 \\
    Cyclone DDS & 717 & 883 & 15 & 4 & 11 & 389 & 215 & 174 & 1,196 & 498 & 698 \\
    Ditto & 34,312 & 140,737 & 157 & 71 & 86 & 47,304 & 10,344 & 36,960 & 127,588 & 23,897 & 103,691 \\
    Embedded CDT & 1,468 & 3,336 & 4,320 & 1,103 & 3,217 & 76 & 57 & 19 & 408 & 308 & 100 \\
    Hawkbit & 196 & 1,462 & 23 & 3 & 20 & 723 & 121 & 602 & 912 & 72 & 840 \\
    Hono & 196 & 879 & 20 & 7 & 13 & 309 & 102 & 207 & 746 & 87 & 659 \\
    Kapua & 3,427 & 16,813 & 480 & 65 & 415 & 9,197 & 1,675 & 7,522 & 10,563 & 1,687 & 8,876 \\
    Kura & 3,802 & 8,626 & 679 & 100 & 579 & 3,694 & 1,708 & 1,986 & 8,055 & 1,994 & 6,061 \\
    Milo & 840,035 & 245,469 & 322 & 240 & 82 & 690,159 & 444,772 & 245,387 & 395,023 & 395,023 & 0 \\
    Mosquitto & 92 & 328 & 16 & 1 & 15 & 100 & 31 & 69 & 304 & 60 & 244 \\
    Mraa & 97 & 285 & 2 & 0 & 2 & 114 & 34 & 80 & 266 & 63 & 203 \\
    Paho & 304 & 480 & 199 & 43 & 156 & 180 & 92 & 88 & 405 & 169 & 236 \\
    VOLTTRON & 81 & 64 & 16 & 2 & 14 & 44 & 32 & 12 & 85 & 47 & 38 \\
    Vorto & 371 & 637 & 82 & 20 & 62 & 186 & 92 & 94 & 740 & 259 & 481 \\ \hline
    \end{tabular}
    }
\end{table*}

\begin{table*}[h!]
    \centering
    \caption{Percentage Distribution of Clone Pairs over Clone Patterns (RQ2)}
    \label{table:PercentClonePairsDistribution}
    \scalebox{0.95}{
    \begin{tabular}{c|ll|cc|cc|cc}
\hline
\multirow{2}{*}{\textbf{Project}} & \multicolumn{2}{c|}{\textbf{Total}} & \multicolumn{2}{c|}{\textbf{Type-1}} & \multicolumn{2}{c|}{\textbf{Type-2}} & \multicolumn{2}{c}{\textbf{Type-3}} \\ \cline{2-9} 
 & \multicolumn{1}{c}{\textbf{IaC}} & \multicolumn{1}{c|}{\textbf{IeC}} & \textbf{IaC} & \textbf{IeC} & \textbf{IaC} & \textbf{IeC} & \textbf{IaC} & \textbf{IeC} \\ \hline
Arrowhead & 11.11\% & 88.89\% & 9.58\% & 90.42\% & 24.81\% & 75.19\% & 11.56\% & 88.44\% \\
Californium & 36.56\% & 63.44\% & 25.64\% & 74.36\% & 55.70\% & 44.30\% & 31.67\% & 68.33\% \\
Cyclone DDS & 44.81\% & 55.19\% & 26.67\% & 73.33\% & 55.27\% & 44.73\% & 41.64\% & 58.36\% \\
Ditto & 19.60\% & 80.40\% & 45.22\% & 54.78\% & 21.87\% & 78.13\% & 18.73\% & 81.27\% \\
Embedded CDT & 30.56\% & 69.44\% & 25.53\% & 74.47\% & 75.00\% & 25.00\% & 75.49\% & 24.51\% \\
Hawkbit & 11.82\% & 88.18\% & 13.04\% & 86.96\% & 16.74\% & 83.26\% & 7.89\% & 92.11\% \\
Hono & 18.23\% & 81.77\% & 35.00\% & 65.00\% & 33.01\% & 66.99\% & 11.66\% & 88.34\% \\
Kapua & 16.93\% & 83.07\% & 13.54\% & 86.46\% & 18.21\% & 81.79\% & 15.97\% & 84.03\% \\
Kura & 30.59\% & 69.41\% & 14.73\% & 85.27\% & 46.24\% & 53.76\% & 24.75\% & 75.25\% \\
Milo & 77.39\% & 22.61\% & 74.53\% & 25.47\% & 64.44\% & 35.56\% & 100.00\% & 0.00\% \\
Mosquitto & 21.90\% & 78.10\% & 6.25\% & 93.75\% & 31.00\% & 69.00\% & 19.74\% & 80.26\% \\
Mraa & 25.39\% & 74.61\% & 0.00\% & 100.00\% & 29.82\% & 70.18\% & 23.68\% & 76.32\% \\
Paho & 38.78\% & 61.22\% & 21.61\% & 78.39\% & 51.11\% & 48.89\% & 41.73\% & 58.27\% \\
VOLTTRON & 55.86\% & 44.14\% & 12.50\% & 87.50\% & 72.73\% & 27.27\% & 55.29\% & 44.71\% \\
Vorto & 36.81\% & 63.19\% & 24.39\% & 75.61\% & 49.46\% & 50.54\% & 35.00\% & 65.00\% \\ \hline
\textbf{Ave} & 31.76\% & 68.24\% & 23.22\% & 76.78\% & 43.03\% & 56.97\% & 34.32\% & 65.68\% \\ \hline
\end{tabular}}
\end{table*}

\begin{tcolorbox}
\textbf{Finding 2}: The distribution of code clones between different commits, whether as a whole or broken down into three types of clones, predominantly consists of inter-commit clones in most projects. However, it is noteworthy that there are still a considerable number of intra-commit clones.
\end{tcolorbox}

\subsection{Evolutionary Trends of Code Clones (RQ3)}
To answer RQ3, we ran a code clone detection across six continuous release versions for each project and deduplicate them as described in Step 1 of Section \ref{Study Procedures}, obtaining the code clone results for those versions, as shown in Table \ref{table:codeClonesDifferentVersions}. The \textit{Latest} column is the latest release version of the project, and the \textit{Rx} column represents the \textit{xth} version prior the latest release version. The \textit{Trend} column shows the trend graph of these data.

Visual inspection of Table \ref{table:codeClonesDifferentVersions} suggests varying evolutionary patterns. To rigorously quantify the evolutionary trends and move beyond visual inspection, we applied the Spearman's rank correlation test (Spearman's $\rho$) as described in Section \ref{sec:dataAnalysis}.

The statistical results (as appended in Table \ref{table:codeClonesDifferentVersions}) reveal three distinct evolutionary patterns among the 15 projects:
(1) Significant Increase (6 projects): Projects such as \texttt{Kapua} ($\rho=1.00$), \texttt{Arrowhead} ($\rho=0.97$), \texttt{Kura} ($\rho=0.94$), and \texttt{Mosquitto} ($\rho=0.93$) show a strong, statistically significant upward trend ($p\textless{0.05}$). This confirms that as these active projects evolve, technical debt in the form of code clones is continuously accumulating. 
(2) Stable Equilibrium (8 projects): The majority of projects, including \texttt{Cyclone DDS}, \texttt{Milo}, \texttt{Embedded CDT}, \texttt{Hawkbit}, \texttt{Hono}, \texttt{Paho}, \texttt{Vorto} and \texttt{Ditto}, show no statistically significant trend ($p\geq0.05$). For example, despite \texttt{Milo} having over 1 million clone pairs, the number remains remarkably constant ($\rho=0.39,p=0.44$), indicating that clone management is balanced. 
(3) Significant Decrease (1 project): Only \texttt{VOLTTRON} exhibits a significant decreasing trend ($\rho=-0.94,p=0.005$), driven by a sharp reduction in recent versions.

\begin{table*}[h!]
    \centering
    \caption{Code Clones in Different Versions (RQ3)}
    \label{table:codeClonesDifferentVersions}
    \scalebox{1.0}{
    \begin{tabular}{c c c c c c c c c c c c c}
        \toprule
        \textbf{Project} & \textbf{R5} & \textbf{R4} & \textbf{R3} & \textbf{R2}  & \textbf{R1} & \textbf{Latest} & \textbf{Trend} & \textbf{$\rho$} & \textbf{p-value}\\
        \midrule
        Arrowhead & 3,966 & 3,966 & 6,678 & 27,220 & 27,220 & 27,223 &
        \begin{tikzpicture}
            \begin{axis}[
                hide axis,
                width=1cm,
                height=0.5cm,
                scale only axis,
                domain=1:6,
                samples=6,
                ytick=\empty,
                xtick={1,2,3,4,5,6},
                xticklabels={1,2,3,4,5,6},
                axis lines=none
            ]
            \addplot[blue, thick] coordinates {(1,3966) (2,3966) (3,6678) (4,27220) (5,27220) (6,27223)};
            \end{axis}
        \end{tikzpicture} 
        & 0.97 & 0.001 \\
        Californium & 975 & 978 & 978 & 981 & 980 & 1,113 & 
        \begin{tikzpicture}
            \begin{axis}[
                hide axis,
                width=1cm,
                height=0.5cm,
                scale only axis,
                domain=1:6,
                samples=6,
                ytick=\empty,
                xtick={1,2,3,4,5,6},
                xticklabels={1,2,3,4,5,6},
                axis lines=none,
                ymin=800,
                ymax=1120
            ]
            \addplot[blue, thick] coordinates {(1,975) (2,978) (3,978) (4,981) (5,980) (6,1113)};
            \end{axis}
        \end{tikzpicture}
        & 0.93 & 0.008 \\
        Cyclone DDS & 2,282 & 2,282 & 2,282 & 2,282 & 2,312 & 1,600 & 
        \begin{tikzpicture}
            \begin{axis}[
                hide axis,
                width=1cm,
                height=0.5cm,
                scale only axis,
                domain=1:6,
                samples=6,
                ytick=\empty,
                xtick={1,2,3,4,5,6},
                xticklabels={1,2,3,4,5,6},
                axis lines=none
            ]
            \addplot[blue, thick] coordinates {(1,2282) (2,2282) (3,2282) (4,2282) (5,2312) (6,1600)};
            \end{axis}
        \end{tikzpicture}
        & -0.17 & 0.749 \\
        Ditto & 174,500 & 174,500 & 174,720 & 174,713 & 174,715 & 174,494 & 
        \begin{tikzpicture}
            \begin{axis}[
                hide axis,
                width=1cm,
                height=0.5cm,
                scale only axis,
                domain=1:6,
                samples=6,
                ytick=\empty,
                xtick={1,2,3,4,5,6},
                xticklabels={1,2,3,4,5,6},
                axis lines=none,
                ymin=174000,  % 扩大 y 轴的最小值
                ymax=175000   % 扩大 y 轴的最大值
            ]
            \addplot[blue, thick] coordinates {(1,174500) (2,174500) (3,174720) (4,174713) (5,174715) (6,174494)};
            \end{axis}
        \end{tikzpicture}
        & -0.06 & 0.913 \\
        Embedded CDT & 4,893 & 4,019 & 4,019 & 4,893 & 4,893 & 4,804 &
        \begin{tikzpicture}
            \begin{axis}[
                hide axis,
                width=1cm,
                height=0.5cm,
                scale only axis,
                domain=1:6,
                samples=6,
                ytick=\empty,
                xtick={1,2,3,4,5,6},
                xticklabels={1,2,3,4,5,6},
                axis lines=none
            ]
            \addplot[blue, thick] coordinates {(1,4893) (2,4019) (3,4019) (4,4893) (5,4893) (6,4804)};
            \end{axis}
        \end{tikzpicture}
        & 0.12 & 0.816 \\
        Hawkbit & 3,198 & 3,853 & 3,782 & 3,804 & 3,795 & 1,657 &
        \begin{tikzpicture}
            \begin{axis}[
                hide axis,
                width=1cm,
                height=0.5cm,
                scale only axis,
                domain=1:6,
                samples=6,
                ytick=\empty,
                xtick={1,2,3,4,5,6},
                xticklabels={1,2,3,4,5,6},
                axis lines=none
            ]
            \addplot[blue, thick] coordinates {(1,3198) (2,3853) (3,3782) (4,3804) (5,3795) (6,1657)};
            \end{axis}
        \end{tikzpicture}
        & -0.26 & 0.623 \\
        Hono & 1,020 & 1,044 & 1,086 & 1,076 & 1,089 & 1,075 & 
        \begin{tikzpicture}
            \begin{axis}[
                hide axis,
                width=1cm,
                height=0.5cm,
                scale only axis,
                domain=1:6,
                samples=6,
                ytick=\empty,
                xtick={1,2,3,4,5,6},
                xticklabels={1,2,3,4,5,6},
                axis lines=none
            ]
            \addplot[blue, thick] coordinates {(1,1020) (2,1044) (3,1086) (4,1076) (5,1089) (6,1075)};
            \end{axis}
        \end{tikzpicture}
        & 0.60 & 0.208 \\
        Kapua & 13,818 & 14,191 & 14,480 & 14,505 & 18,021 & 20,013 & 
        \begin{tikzpicture}
            \begin{axis}[
                hide axis,
                width=1cm,
                height=0.5cm,
                scale only axis,
                domain=1:6,
                samples=6,
                ytick=\empty,
                xtick={1,2,3,4,5,6},
                xticklabels={1,2,3,4,5,6},
                axis lines=none
            ]
            \addplot[blue, thick] coordinates {(1,13818) (2,14191) (3,14480) (4,14505) (5,18021) (6,20013)};
            \end{axis}
        \end{tikzpicture}
        & 1.00 & 0.000 \\
        Kura  & 9,297 & 9,322 & 11,308 & 11,308 & 11,308 & 11,952 & 
        \begin{tikzpicture}
            \begin{axis}[
                hide axis,
                width=1cm,
                height=0.5cm,
                scale only axis,
                domain=1:6,
                samples=6,
                ytick=\empty,
                xtick={1,2,3,4,5,6},
                xticklabels={1,2,3,4,5,6},
                axis lines=none
            ]
            \addplot[blue, thick] coordinates {(1,9297) (2,9322) (3,11308) (4,11308) (5,11308) (6,11952)};
            \end{axis}
        \end{tikzpicture}
        & 0.94 & 0.005 \\
        Milo & 1,085,502 & 1,085,502 & 1,085,502 & 1,085,502 & 1,085,503 & 1,085,502 & 
        \begin{tikzpicture}
            \begin{axis}[
                hide axis,
                width=1cm,
                height=0.5cm,   
                scale only axis,
                domain=1:6,
                samples=6,
                ytick=\empty,
                xtick={1,2,3,4,5,6},
                xticklabels={1,2,3,4,5,6},
                axis lines=none
            ]
            \addplot[blue, thick] coordinates {(1,1085502) (2,1085502) (3,1085502) (4,1085502) (5,1085503) (6,1085502)};
            \end{axis}
        \end{tikzpicture}
        & 0.39 & 0.441 \\
        Mosquitto & 382 & 382 & 412 & 419 & 419 & 419 &
        \begin{tikzpicture}
            \begin{axis}[
                hide axis,
                width=1cm,
                height=0.5cm,
                scale only axis,
                domain=1:6,
                samples=6,
                ytick=\empty,
                xtick={1,2,3,4,5,6},
                xticklabels={1,2,3,4,5,6},
                axis lines=none
            ]
            \addplot[blue, thick] coordinates {(1,382) (2,382) (3,412) (4,419) (5,419) (6,419)};
            \end{axis}
        \end{tikzpicture}
        & 0.93 & 0.008 \\
        Mraa & 188 & 253 & 288 & 287 & 317 & 382 &
        \begin{tikzpicture}
            \begin{axis}[
                hide axis,
                width=1cm,
                height=0.5cm,
                scale only axis,
                domain=1:6,
                samples=6,
                ytick=\empty,
                xtick={1,2,3,4,5,6},
                xticklabels={1,2,3,4,5,6},
                axis lines=none
            ]
            \addplot[blue, thick] coordinates {(1,188) (2,253) (3,288) (4,287) (5,317) (6,382)};
            \end{axis}
        \end{tikzpicture}
        & 0.94 & 0.005 \\
        Paho & 392 & 799 & 801 & 800 & 800 & 766 &
        \begin{tikzpicture}
            \begin{axis}[
                hide axis,
                width=1cm,
                height=0.5cm,
                scale only axis,
                domain=1:6,
                samples=6,
                ytick=\empty,
                xtick={1,2,3,4,5,6},
                xticklabels={1,2,3,4,5,6},
                axis lines=none
            ]
            \addplot[blue, thick] coordinates {(1,392) (2,799) (3,801) (4,800) (5,800) (6,766)};
            \end{axis}
        \end{tikzpicture}
        & 0.23 & 0.658 \\
        VOLTTRON & 466,066 & 466,066 & 466,066 & 246,553 & 389 & 145 & 
        \begin{tikzpicture}
            \begin{axis}[
                hide axis,
                width=1cm,
                height=0.5cm,
                scale only axis,
                domain=1:6,
                samples=6,
                ytick=\empty,
                xtick={1,2,3,4,5,6},
                xticklabels={1,2,3,4,5,6},
                axis lines=none
            ]
            \addplot[blue, thick] coordinates {(1,466066) (2,466066) (3,466066) (4,246553) (5,389) (6,145)};
            \end{axis}
        \end{tikzpicture}
        & -0.94 & 0.005 \\
        Vorto & 1,023 & 1,024 & 1,024 & 1,023 & 1,023 & 1,008  &
        \begin{tikzpicture}
            \begin{axis}[
                hide axis,
                width=1cm,
                height=0.5cm,
                scale only axis,
                domain=1:6,
                samples=6,
                ytick=\empty,
                xtick={1,2,3,4,5,6},
                xticklabels={1,2,3,4,5,6},
                axis lines=none,
                ymin=940,  % 扩大 y 轴的最小值
                ymax=1040   % 扩大 y 轴的最大值
            ]
            \addplot[blue, thick] coordinates {(1,1023) (2,1024) (3,1024) (4,1023) (5,1023) (6,1008)};
            \end{axis}
        \end{tikzpicture}
        & -0.62 & 0.192 \\
        \bottomrule
    \end{tabular}
    }
\end{table*}

\begin{tcolorbox}
    \textbf{Finding 3}: Confirmed by Spearman's rank correlation analysis, 14 out of 15 projects (93\%) exhibit either a stable or significantly increasing trend in code clones. Specifically, 6 active projects show a strong accumulation of clones ($\rho>0.9,p<0.05$), while 8 projects maintain a dynamic equilibrium. Significant reduction in code clones is exceptionally rare, observed only in one project.
\end{tcolorbox}

\subsection{Co-modified Code Clones (RQ4)}
To answer RQ4, we performed code clone detection between all versions of the project as described in Step 3 of Section \ref{Study Procedures}, and deduplicated the results to obtain co-modified clone pairs for each project. We further analyzed these clone pairs to identify whether they are intra-commit or inter-commit clones.

The co-modified code clone detection results are presented in Table \ref{table:coModifiedCloneAnalysisIIoT}, in which the \textit{CoPairs} column denotes the number of co-modified clone pairs, the \textit{Percentage} column represents the proportion of co-modified clone pairs to the total number of clone pairs, the \textit{Tx Pairs} column denotes clone type, and the \textit{IaC Pairs} and \textit{IeC Pairs} columns denote the number of intra-commit clones and the number of inter-commit clones, respectively.
\begin{table*}[h!]
    \centering
    \caption{Results of Co-modified Clone Analysis for Eclipse IIoT OSS Projects (RQ4)}
    \label{table:coModifiedCloneAnalysisIIoT}
    \scalebox{1.0}{
    \begin{tabular}{c c c c c c c c}
        \toprule
        \textbf{Project} & \textbf{CoPairs} & \textbf{Percentage} & \textbf{Type-1 Pairs} & \textbf{Type-2 Pairs} & \textbf{Type-3 Pairs} & \textbf{IaC Pairs} & \textbf{IeC Pairs} \\
        \midrule
        Arrowhead & 0 & 0.00\% & 0 & 0 & 0 & 0 & 0 \\
        Californium & 4 & 0.07\% & 0 & 1 & 3 & 3 & 1 \\
        Cyclone DDS & 267 & 5.20\% & 0 & 49 & 218 & 132 & 135 \\
        Ditto & 5 & 0.003\% & 0 & 0 & 5 & 2 & 3 \\
        Embedded CDT & 0 & 0.00\% & 0 & 0 & 0 & 0 & 0 \\ 
        Hawkbit & 50 & 1.95\% & 4 & 11 & 35 & 14 & 36 \\
        Hono & 10 & 0.39\% & 0 & 1 & 9 & 3 & 7 \\
        Kapua & 1,505 & 5.16\% & 5 & 575 & 925 & 309 & 1,196 \\
        Kura  & 513 & 1.53\% & 7 & 361 & 145 & 445 & 68 \\
        Milo & 0 & 0.00\% & 0 & 0 & 0 & 0 & 0 \\ 
        Mosquitto & 8 & 0.31\% & 0 & 2 & 6 & 5 & 3 \\
        Mraa & 84 & 20.49\% & 2 & 27 & 55 & 28 & 56 \\
        Paho & 32 & 2.10\% & 1 & 2 & 29 & 30 & 2 \\
        VOLTTRON & 2 & 0.26\% & 0 & 1 & 1 & 0 & 2 \\
        Vorto & 9 & 0.04\% & 0 & 2 & 7 & 5 & 4 \\
        \hline
        Total/Ave & 2,489 & 0.17\% & 19 & 1,032 & 1,438 & 976 & 1,513 \\  
        \bottomrule
    \end{tabular}}
\end{table*}

% \begin{table*}[h!]
%     \centering
%     \caption{Percentage of clone pairs distribution in co-modified clones (RQ4)}
%     \label{table:PercentOfCoModified}
%     \scalebox{0.96}{
%     \begin{tabular}{c c c c c c c c c c c c}
%         \toprule
%         \textbf{All} & \textbf{Percentage} &
%         \textbf{T1} & 
%         \textbf{Percentage} & \textbf{T2} & 
%         \textbf{Percentage} & \textbf{T3} & 
%         \textbf{Percentage} & \textbf{IaC} & 
%         \textbf{Percentage} & \textbf{IeC} &
%         \textbf{Percentage} \\
%         \midrule
%         2,489 & 0.17\% & 19 & 0.76\% & 1,032 & 41.46\% & 1,438 & 57.78\% & 976 & 39.21\% & 1,513 & 60.79\% \\
%         \bottomrule
%     \end{tabular}}
% \end{table*}

Based on the data presented in Table \ref{table:coModifiedCloneAnalysisIIoT}, we can obtain several key findings. Firstly, co-modified clones are present in most Eclipse IIoT OSS projects. Secondly, T3 clones are the most likely to occur, constituting 57.78\% ($\frac{1,438}{2,489}$) of the total number of clones. Lastly, more than half of the co-modified clones are inter-commit clones. As the number of T3 clones continues to rise, and given that inter-commit clones also represent a significant proportion, we believe that code maintenance driven by co-modifications will escalate with the growth of project scale, leading to increased maintenance overhead. 

However, the absolute proportion of these co-modifications is relatively small; only three projects have more than 100 co-modifications. It is also worth noting that there exist marked differences among projects: for example, some projects (e.g., \texttt{Milo}) with a large number of code clones may have very few or even no co-modified clones, while others (e.g., \texttt{Mraa}) with relatively few overall clones may exhibit a higher proportion of co-modified clones. Therefore, we plan to further investigate the frequency of co-modifications across different projects and their specific impact on code maintenance, in order to more comprehensively assess and address the growing maintenance costs.

\begin{tcolorbox}
    \textbf{Finding 4}: Co-modified code clones are identified in the majority of IIoT OSS projects, though not at a high frequency, with Type-3 clones occurring more frequently than Type-1 and Type-2 clones. Over half of the co-modified clones are inter-commit clones. Notably, \texttt{Milo}, the project with the highest number of 1.08M clone pairs, has no co-modified clones; in contrast, \texttt{Mraa}, despite having the fewest clones, has the largest proportion of co-modified clones. Additionally, when looking at the overall data, we can see that the quantity of co-modified code clones in these projects is not large, with only three projects exceeding one hundred.
\end{tcolorbox}

\subsection{Cross-project Code Clones (RQ5)}
%To conduct an in-depth study of the maintenance issues that may arise from code clone, 
To answer RQ5, we examined cross-project clones between different projects. Given that Eclipse IIoT is a complex ecosystem, clones between projects may result from cross-project code reuse within the ecosystem. In particular, cross-project clones can lead to version synchronization problems and potential defect propagation, which requires special attention. Therefore, we analyzed the distribution of clone pairs across different projects. The results are presented in Table \ref{table:crossProjectsCloneAnalysis}.

\begin{table*}[]
    \caption{Results of Cross-projects Clone Analysis for IIoT OSS Projects (RQ5)}
    % \scriptsize
    \label{table:crossProjectsCloneAnalysis}
    \scalebox{1.0}{
    \begin{tabular}{ccccccc|cccclcl}
\hline
\multicolumn{7}{c|}{\textbf{Java}}                                                                                                                                                 & \multicolumn{7}{c}{C}                                                                                                                                                                                                              \\ \hline
\multicolumn{1}{c|}{\textbf{Sum}}   & \multicolumn{2}{c|}{\textbf{Type-1}}                    & \multicolumn{2}{c|}{\textbf{Type-2}}                 & \multicolumn{2}{c|}{\textbf{Type-3}} & \multicolumn{1}{c|}{\textbf{Sum}}   & \multicolumn{2}{c|}{\textbf{Type-1}}                    & \multicolumn{2}{c|}{\textbf{Type-2}}                                        & \multicolumn{2}{c}{\textbf{Type-3}}                              \\ \hline
\multicolumn{1}{c|}{\textbf{Pairs}} & \textbf{Pairs} & \multicolumn{1}{c|}{\textbf{CLOC}} & \textbf{Pairs} & \multicolumn{1}{c|}{\textbf{CLOC}} & \textbf{Pairs}  & \textbf{CLOC}  & \multicolumn{1}{c|}{\textbf{Pairs}} & \textbf{Pairs} & \multicolumn{1}{c|}{\textbf{CLOC}} & \multicolumn{1}{l}{\textbf{Pairs}} & \multicolumn{1}{l|}{\textbf{CLOC}} & \multicolumn{1}{l}{\textbf{Pairs}} & \textbf{CLOC}           \\ \hline
\multicolumn{1}{c|}{73,264}         & 41             & \multicolumn{1}{c|}{812}           & 2,911          & \multicolumn{1}{c|}{5,920}         & 70,312          & 33,661         & \multicolumn{1}{c|}{74}             & 0              & \multicolumn{1}{c|}{0}             & 2                                  & \multicolumn{1}{c|}{22}            & 72                                 & \multicolumn{1}{c}{766} \\ \hline
\end{tabular}
    }
\end{table*}

Table \ref{table:crossProjectsCloneAnalysis} shows that source code in both Java and C has cross-project clones, with Java code having significantly more cross-project clones than C code. Regardless of whether the code is in Java or C, the most common clone type is Type-3, which aligns with our Finding 1, as Type-3 clones are the most frequent in both co-modified and ordinary clones. These results also highlight the existence of cross-project clones in Eclipse IIoT OSS projects. Such cross-project clones cannot only increase maintenance costs but also elevate the risk of error propagation between different projects. Changes made to a code block may be propagated to all other clone blocks of that block, potentially leading to additional challenges in maintaining software integrity. 

To further uncover the existence of these cross-project clones, we extracted and sorted these clone pairs, resulting in the top 10 for different clone types, as shown in Table \ref{table:top10CrossProjectClonePairs}. We can observe that in Java projects, \texttt{Kura} and \texttt{Kapua} contain the most cross-project code clones, while in C projects, \texttt{Mraa} and \texttt{Cyclone DDS} are the ones in which the most cross-project code clones occur. 

\begin{table*}[]
    \caption{Cross-project Clone Pairs Top-10 (RQ5)}
    % \scriptsize
    \label{table:top10CrossProjectClonePairs}
    \scalebox{1.0}{
    \begin{tabular}{lllcccccc}
    \hline
    \multicolumn{9}{c}{\textbf{Java}}                                                             \\ \hline
    \multicolumn{3}{c|}{\textbf{Type-1}}                                                                                      & \multicolumn{3}{c|}{\textbf{Type-2}}                                              & \multicolumn{3}{c}{\textbf{Type-3}}                          \\ \hline
    \multicolumn{1}{c}{\textbf{Project 1}} & \multicolumn{1}{c}{\textbf{Project 2}} & \multicolumn{1}{c|}{\textbf{Pairs}} & \textbf{Project 1} & \textbf{Project 2} & \multicolumn{1}{c|}{\textbf{Pairs}} & \textbf{Project 1} & \textbf{Project 2} & \textbf{Pairs} \\ \hline
    \multicolumn{1}{c}{Kapua}              & \multicolumn{1}{c}{Kura}               & \multicolumn{1}{c|}{41}             & Kapua              & Milo               & \multicolumn{1}{c|}{563}            & Ditto              & Arrowhead          & 9,430           \\
                                           &                                        & \multicolumn{1}{l|}{}               & Kapua              & Kura               & \multicolumn{1}{c|}{407}            & Arrowhead          & Milo               & 6,682           \\
                                           &                                        & \multicolumn{1}{l|}{}               & Arrowhead          & Hawkbit            & \multicolumn{1}{c|}{337}            & Kura               & Milo               & 4,268           \\
                                           &                                        & \multicolumn{1}{l|}{}               & Kura               & Milo               & \multicolumn{1}{c|}{301}            & Kura               & Arrowhead          & 4,219           \\
                                           &                                        & \multicolumn{1}{l|}{}               & Ditto              & Arrowhead          & \multicolumn{1}{c|}{254}            & Kapua              & Kura               & 3,839           \\
                                           &                                        & \multicolumn{1}{l|}{}               & Vorto              & Milo               & \multicolumn{1}{c|}{150}            & Kapua              & Arrowhead          & 3,468           \\
                                           &                                        & \multicolumn{1}{l|}{}               & Kapua              & Vorto              & \multicolumn{1}{c|}{138}            & Kapua              & Milo               & 3,460           \\
                                           &                                        & \multicolumn{1}{l|}{}               & Vorto              & Kura               & \multicolumn{1}{c|}{126}            & Ditto              & Milo               & 2,960           \\
                                           &                                        & \multicolumn{1}{l|}{}               & Kura               & Arrowhead          & \multicolumn{1}{c|}{118}            & Ditto              & Hono               & 2,760           \\
                                           &                                        & \multicolumn{1}{l|}{}               & Kapua              & Arrowhead          & \multicolumn{1}{c|}{101}            & Ditto              & Kura               & 2,664           \\ \hline
    \multicolumn{9}{c}{\textbf{C}}                                                                                                                                                                                                                                   \\ \hline
    \multicolumn{3}{c|}{\textbf{Type-1}}                                                                                      & \multicolumn{3}{c|}{\textbf{Type-2}}                                              & \multicolumn{3}{c}{\textbf{Type-3}}                          \\ \hline
    \multicolumn{1}{c}{\textbf{Project 1}} & \multicolumn{1}{c}{\textbf{Project 2}} & \multicolumn{1}{c|}{\textbf{Pairs}} & \textbf{Project 1} & \textbf{Project 2} & \multicolumn{1}{c|}{\textbf{Pairs}} & \textbf{Project 1} & \textbf{Project 2} & \textbf{Pairs} \\ \hline
     None                                      &     None                                   & \multicolumn{1}{c|}{0}               & Mraa               & Cyclone DDS        & \multicolumn{1}{c|}{2}              & Mraa               & Cyclone DDS        & 44             \\
                                           &                                        & \multicolumn{1}{l|}{}               &                    &                    & \multicolumn{1}{c|}{}               & Cyclone DDS        & Mosquitto          & 25             \\
                                           &                                        & \multicolumn{1}{l|}{}               &                    &                    & \multicolumn{1}{c|}{}               & Mraa               & Mosquitto          & 3              \\ \hline
    \end{tabular}
    }
\end{table*}

To further investigate the characteristic changes of these cross-project code clones across versions, we collected the last five release versions of all projects and examined the likelihood of co-modifications among these projects. Ultimately, we identified 14 pairs of co-modified clones among these cross-project clones. After removing duplicates, 11 unique pairs remained. The results are presented in Table \ref{table:coModifiedCrossProjectCodeClones}, in which \textit{<x, y>} represents the two projects in the co-modified clone pair. The results indicate that there are indeed instances of co-modifications among these cross-project code clones; however, the number of such occurrences is very small, resulting in a negligible impact on maintenance, as only 11 pairs were detected in the Java language, accounting for just 0.02\% (11/73,264) of all cross-project code clones.

\begin{table}[]
\caption{Co-modification in Cross-project Code Clones (RQ5)}
    % \scriptsize
    \label{table:coModifiedCrossProjectCodeClones}
    \scalebox{1.0}{
   \begin{tabular}{cc|cc}
\hline
\multicolumn{2}{c|}{\textbf{Java}}                                 & \multicolumn{2}{c}{\textbf{C}}                        \\ \hline
\textbf{CoPairs}     & \textbf{Detail}                             & \textbf{CoPairs}    & \textbf{Detail}                 \\ \hline
\multirow{11}{*}{11} & \textless{}Vorto, Arrowhead\textgreater{}   & \multirow{11}{*}{0} & \multirow{11}{*}{\textbf{None}} \\
                     & \textless{}Hono, Vorto\textgreater{}        &                     &                                 \\
                     & \textless{}Vorto, Californium\textgreater{} &                     &                                 \\
                     & \textless{}Kura, Kapua\textgreater{}        &                     &                                 \\
                     & \textless{}Hawkbit, Arrowhead\textgreater{} &                     &                                 \\
                     & \textless{}Hawkbit, Kura\textgreater{}      &                     &                                 \\
                     & \textless{}Arrowhead, Hawkbit\textgreater{} &                     &                                 \\
                     & \textless{}Arrowhead, Ditto\textgreater{}   &                     &                                 \\
                     & \textless{}Vorto, Hawkbit\textgreater{}     &                     &                                 \\
                     & \textless{}Vorto, Californium\textgreater{} &                     &                                 \\
                     & \textless{}Kura, Vorto\textgreater{}        &                     &                                 \\ \hline
\end{tabular}
    }
\end{table}

\begin{tcolorbox}
    \textbf{Finding 5}: Cross-project code clones are prevalent in the Eclipse IIoT OSS project. Notably, the prevalence of cross-project code clones in Java projects is significantly higher than that in C projects. Specifically, projects \texttt{Kura} and \texttt{Kapua} exhibit a high density of cross-project code clones within Java, while projects \texttt{Mraa} and \texttt{Cyclone DDS} dominate in C projects. However, these cross-project code clones have a co-modification rate of only 0.02\%, with all occurrences found in Java projects.
\end{tcolorbox}

\section{Discussion}\label{sec_discussion}
In this section, we first interpret the study results, and then discuss their implications. %We aim to extract lessons from them and provide suitable recommendations for the development, maintenance, and use of future IIoT OSS projects. In addition, we intend to guide future research on code clones in related software.

\subsection{Understanding the Results}\label{understandResults}
\textbf{RQ1:} \textbf{Finding 1 confirms the presence and prevalence of code clones in the Eclipse IIoT OSS projects.} With cloned code lines accounting for $16.30\%$ of the total codebase on average. To establish a concrete baseline for understanding the prevalence of code clones, we contextualize our findings against several emerging and traditional software domains. Our results show that the prevalence of code clones in IIoT is significantly higher than in specialized domains such as Solidity Smart Contracts \cite{Mo2025CodeCloneOnSSC} and Virtual Reality (VR) software \cite{Huang2024CodeCloneOSVRS}. 

In smart contracts, while code clones are common at the function level, the actual proportion of cloned lines is only $5.12\%$ on average (ranging from $2.78\%$ to $21.61\%$) \cite{Mo2025CodeCloneOnSSC}. Similarly, in VR software, the source code clone ratio remains low at approximately $5.02\%$ on average, with high redundancy instead occurring in serialized asset files ($43.47\%$ to $91.63\%$) \cite{Huang2024CodeCloneOSVRS}. In contrast, our study finds that redundancy in IIoT software is more deeply embedded within the logic-intensive source code, specifically in the implementation of industrial protocols and device configurations. Furthermore, our results indicate that the clone density in IIoT projects is nearly twice that of traditional OSS projects, which typically exhibit a benchmark clone ratio of approximately 8.60\% \cite{Mo2023ExploringTI}. This prevalence is notably higher than the levels observed in large-scale C/C++ systems, which report an average clone ratio of approximately 12\% \cite{Koschke2016Saner}, and it aligns with the upper tier of the general range for traditional software (including Java), which typically falls between 7\% and 23\% \cite{Jebnoun2021ClonesID}. Notably, the clone intensity of IIoT software is remarkably consistent with the levels observed in other specialized domains, such as deep learning software. For instance, empirical evidence shows that deep learning software exhibits an identical average clone proportion of 16.3\% \cite{Mo2023ExploringTI}. This suggests that IIoT software is as clone-intensive as deep learning software, likely caused by the fundamental necessity for adaptive reuse, i.e., where logic is frequently copied and modified to accommodate heterogeneous devices and evolving protocol constraints.

However, the prevalence of code clone in IIoT software has not yet reached the extreme levels found in high-stakes safety-critical systems like autonomous driving software. In systems such as Apollo and Autoware, the average proportion of cloned lines ($PLOC$) ranges from $31.6\%$ to $35.3\%$, with some versions peaking at $38.8\%$ \cite{Mo2023ACS}. This comparative spectrum indicates that the prevalence of code clones in the Eclipse IIoT ecosystem is situated at a middle-high tier, i.e., surpassing smart contracts and VR code, matching deep learning frameworks, but remaining below the maximal redundancy seen in the automotive domain.

On the other hand, Finding 1 reveals the notable difference in the proportions of cloned code for different projects. During the analysis, we found that the number of clones in the \texttt{Milo} and \texttt{Ditto} projects is relatively high, whereas \texttt{Mraa} and \texttt{VOLTTRON} exhibit a comparatively low number of clones. Further investigation revealed that \texttt{Milo} is an open-source implementation of the Open Platform Communications Unified Architecture (OPC UA), while \texttt{Ditto} pertains to software pattern implementations for building digital twins of devices connected to the Internet. Unlike IIoT OSS projects that provide services and interfaces, these framework and pattern implementations often result in extensive code reuse due to the similarity of certain implementation codes, leading to repeated code cloning. For instance, in \texttt{Milo}, there are 1,279 node implementations for loading, resulting in 1,279 highly similar loading functions, which generate over 500K lines of code clones.

In projects with fewer clones, we took \texttt{Mraa} and \texttt{VOLTTRON} as representatives. \texttt{Mraa} is a C/C++ library with bindings for Java, Python, and JavaScript, designed to connect I/O pins and buses on various IoT and edge platforms. Because the interfaces for the same language are responsible for different functionalities, and due to the relatively lower modularity and fewer shared functions in C, the notable differences in the implementation of interfaces for different languages contribute to a reduction in code clones. \texttt{VOLTTRON} is primarily written in Python. We found that its development team is currently working on migrating the agent (a stand-alone software component that is responsible for performing a specific task or function, and can run independently and interact with other agents or systems via \texttt{VOLTTRON's} message buses) from a monolithic version to a modular version. This means that the \texttt{VOLTTRON} project mostly consists of monolithic code at present, which results in lower modularity. Since our clone detection operates at the function level, the number of clones detected in monolithic code tends to be relatively low.

\textbf{RQ2:} 
% 更新对提交间克隆和提交内克隆的结果说明
\textbf{Finding 2 reveals a clear temporal split in how code clones occur.} Inter-commit clones reflect ``evolutionary'' reuse, where developers reimplement or copy logic across tasks and over time. These clones often indicate architectural or API-level abstractions that were never factored out. Intra-commit clones capture ``ad-hoc'' reuse within the same coding session — copy–paste bursts used to accelerate feature delivery or bug fixes.
For example, in project \texttt{Milo}, we detected 840,035 intra-commit clone pairs, compared to only 245,469 inter-commit pairs. In fact, in this project, most intra-commit clones were concentrated in node-loading code (i.e., defining new type nodes in the OPC UA server), where developers repeatedly implemented highly similar routines with identical functionality. Such an imbalance between intra-commit clones and inter-commit clones was most pronounced in this project. This imbalance highlights that, during active development (i.e., periods of frequent code changes and feature additions), local duplication is a deliberate trade-off for speed, but if left unchecked, it can incur technical debt immediately. 

In contrast, project \texttt{Mraa} demonstrates a different pattern, where T1 clones are exclusively inter-commit. Further analysis reveals that identical memory deallocation routines are duplicated across 16 code blocks in 15 separate files, with each instance introduced in different commits over time. This pattern exemplifies evolutionary code reuse, where functionally identical code blocks are repeatedly introduced across the codebase without consolidation, suggesting the need for systematic refactoring to extract common utilities or establish shared libraries.

These findings reveal distinct intervention opportunities for different clone patterns. For intra-commit clones, as demonstrated in the \texttt{Milo} project, the concentration of similar code within single commits presents immediate refactoring opportunities. If code refactoring could be performed before code submissions, through precommit hooks or real-time IDE suggestions, the proportion of such clones could be substantially reduced. The presence of substantial intra-commit code clones also reflects a potential lack of awareness among developers regarding timely refactoring practices during active coding sessions. For inter-commit clones, as seen in the \texttt{Mraa} project, the challenge lies in identifying and consolidating evolutionary duplications that accumulate over time, requiring different detection and remediation strategies.

\textbf{RQ3:} \textbf{Finding 3 illustrates the evolution of code clones during version iterations. In most Eclipse IIoT OSS projects, there has not been a marked decrease as the project size expands and the versions evolve.} Although we can observe some projects showing a decrease of code clones during version iterations, the decrease is not substantial. Furthermore, the number of clones often rebounds to initial levels or even higher in subsequent iterations. This indicates that, overall, the number of code clones in IIoT OSS projects remains relatively stable. This phenomenon suggests that due to factors such as loose coupling, developers may choose not to refactor existing functionalities to reduce code clones but rather replicate these functionalities, leading to an increase in code clones. 

The large number of node loading functions in the \texttt{Milo} project serves as a striking example of this phenomenon. Additionally, in the \texttt{Kura} project, which is used for communication and platform integration, there are numerous adapter and configuration functions, such as \textit{I2CDeviceConfig} for I/O transmission and \textit{PulseCounterConfig} for configuring pulse counters. Most of these configuration functions have similar parameters, with each configurator belonging to a different module. As the number of platform-adaptive interfaces increases, so do the modules, leading to a corresponding increase in configuration functions and, consequently, a rise in code clones.

We identified an anomalous project, \texttt{VOLTTRON}, where we observed that the number of clones exhibited an exponential decline during the last four iterations, with a maximum of 466K and a minimum of 145. \texttt{VOLTTRON} is an open-source distributed sensing and control platform primarily implemented in Python. In Version 8.2, it upgraded to the latest \textit{RabbitMQ (3.9.7)} and \textit{Pika (1.2.0)}, but it is not backward compatible. To ensure that the systems could continue functioning, the original protocol code had to be retained, resulting in high similarity between the two different protocols, which greatly increased the number of clones. As versions iterated and updates occurred, the code for the old protocol was gradually removed, leading to a substantial decrease in code clones.

\textbf{RQ4:} \textbf{Finding 4 validates the existence of co-modified code clones within Eclipse IIoT OSS projects, indicating that code clones indeed impose a considerable burden on project maintenance and modification.} Among the 15 projects under study, 12 have co-modified code clones. Interestingly, although \texttt{Milo} and \texttt{Ditto} have a great number of clone pairs, their instances of co-modified code clones are quite rare. In contrast, \texttt{Mraa} has the fewest clone pairs, but the highest proportion of co-modified code clones. 

Our further investigation revealed that \texttt{Mraa} is an open-source library designed to simplify hardware interactions, particularly in the fields of IoT and embedded development, primarily written in C. For projects that support multiple hardware platforms, developers often need to customize their code based on the specific hardware in use. This leads to a considerable amount of co-modifications required between different versions to accommodate various hardware platforms. On the other hand, open-source protocol implementations like \texttt{Milo} tend to extend existing code rather than modify it extensively. Due to the high reusability and modularity of the existing code as well as its loose coupling, there is minimal occurrence of co-modifications.

When examining the specific numbers of co-modified clones, we find that although 12 projects experienced co-modified code clones, the frequency of the occurrences is relatively low. In fact, three projects have a co-modification rate no more than 0.1\%, six projects have a rate less than 1.0\%, and the overall average co-modification occurrence rate is only 0.17\%. In contrast, previous studies in other domains have shown that 10.3\% of code clones in deep learning software involve co-modifications \citep{Mo2023ExploringTI}, while in autonomous driving software like Apollo and Autoware, the rates are 36.4\% and 11.2\%, respectively \citep{Mo2023ACS}. This suggests that the maintenance challenges posed by co-modifications in Eclipse IIoT OSS projects are much lower.

However, while the detected co-modifications do not seem to complicate maintenance greatly, our analysis of the cloned code reveals that although there are not many changes to existing code, a large amount of newly added code is highly similar to existing code. This similarity can increase the maintenance burden from another perspective. When the project changes, developers must modify many similar functions rather than refactor key functions, which keeps the cost of code changes substantial.This highlights that both the addition of similar code and the maintenance of existing code clones contribute to long-term complexity.

% 增加对深入探讨共更改数量稀少的原因探析
Further analysis suggests that although the rate of co-modifications is low, it remains non-negligible. For instance, in the \texttt{Mosquitto} project, a clone pair identified between versions 2.14.0 and 2.15.0 exhibits a scenario where one part of the clone was modified, while the other was left unchanged. Subsequent updates involved further changes to the unmodified segment, exacerbating maintenance complexity in several ways: developers must trace and understand the relationship between scattered clone blocks, identify which parts require synchronized modifications, and ensure consistency across multiple code locations. This fragmented modification pattern increases the cognitive load on developers, makes debugging more challenging when issues arise, and elevates the risk of introducing inconsistencies due to incomplete or misaligned updates across clone segments. Similarly, in the \texttt{Hono} project, co-modified clones show instances of simultaneous deletions, followed by additions in later updates. These instances are exemplified by frequent changes to the same Java annotations across different versions, demonstrating how clone modifications can become increasingly complex over time. 

These findings indicate that, despite the small proportion of co-modified clones in Eclipse IIoT OSS projects, their impact cannot be overlooked. Developers' lack of awareness of the existence of code clones can lead to uncoordinated modifications, resulting in inconsistencies and increased maintenance costs.

\textbf{RQ5:} \textbf{Finding 5 reveals the prevalence of cross-project code clones in Eclipse IIoT OSS projects, indicating that code reuse across projects is widespread.} However, the co-modification occurrence rate among these cross-project clone pairs is only 0.02\%, meaning that while cross-project co-modified code clones do exist, they are extremely rare. This indicates that, in practice, cross-project code clones in Eclipse IIoT OSS projects does not result in much maintenance overhead.

To explore the reasons for the low incidence of co-modifications in cross-project code clones, we conducted an in-depth study of the functionalities and source code of these projects. We found that: (1) Cross-project code clones primarily occur in the data transmission and communication protocol code between \texttt{Kura}, \texttt{Kapua}, and other projects. Within the same IIoT OSS ecosystem, these implementations are generally similar, leading to a considerable amount of cross-project code clones. (2) Although there are many cross-project code clones, the slower pace of protocol updates and the relative stability of data transmission code compared to other code (such as core functionality code) result in minimal changes to this code, thereby reducing the likelihood of co-modifications between clones. (3) Fundamentally, these projects maintain a degree of independence. Although they use the same communication protocols and data transmission methods, it is not required for these projects to synchronize updates to the relevant code. For example, in cross-project co-modification clones related to communication protocols, if one project upgrades and modifies the communication protocol in its next version while another project does not have the need to update the protocol, the latter does not need to modify the cloned code, thus no co-modification occurs.

In different programming languages, many platform and protocol implementations are primarily coded in Java, and due to the strong modularity of Java, Java projects tend to have notably more instances of both cross-project clones and cross-project co-modifications compared to those written in C. Within Java projects, we found that \texttt{Kura} and \texttt{Kapua} are particularly prone to cross-project code clones with different projects. To uncover the reasons behind this, we conducted an in-depth investigation into these two projects. We found that both \texttt{Kura} and \texttt{Kapua} are designed for data services, while other projects belong to different domains. To focus on specific functionalities, Java projects that are not centered on data collection (such as \texttt{Arrowhead} and \texttt{Milo}) often rely on data collection platforms like \texttt{Kura} and \texttt{Kapua}. We also noted that \texttt{Kura} and \texttt{Kapua} share similar framework implementations and exhibit interdependencies, which make their code more easily reusable by other projects. Similarly, we conducted an in-depth study of C projects \texttt{Mraa} and \texttt{Cyclone DDS}, as these projects demonstrate a higher frequency of cross-project code clones. Our review revealed that this is due to the abundance of library functions in C, which are frequently called by different projects, with \texttt{Mraa} itself being a library. In addition, C projects are mainly used for hardware-related tasks in IIoT, which require more communication protocols. This dependency on protocols leads to increased reuse, contributing to the rise of cross-project code clones between these two projects and others.

\subsection{Comparison with a General Software Study}
To determine whether our observations are specific to IIoT or reflect general software laws, we compared our findings with the large-scale study by Lopes et al. \citep{Lopes2017}, which identified universal clone characteristics across 4.5 million non-fork projects.

(1) Clone Type Divergence (Specific to IIoT): Lopes et al. found that exact copies (Type-1/File-Hash) dominate general software ecosystems (e.g., 94\% in JavaScript, 40\% in Java), caused by the inclusion of entire libraries. In stark contrast, our Finding 1 reveals that Type-3 clones dominate (56.72\%) the Eclipse IIoT software ecosystem. Even accounting for the granularity difference (function vs. file), this suggests a fundamental behavioral difference: IIoT developers engage in ``adaptive reuse'' (i.e., copying logic but modifying it to fit specific hardware or protocol constraints) rather than the ``blind consumption'' of libraries seen in general Web/App development.

(2) The Middle-Size Law (Universal): Lopes et al. observed a ``bell curve`` where clones are most frequent in medium-sized files. We observe a parallel at the function level: clones are concentrated in medium-complexity protocol handlers (e.g., node loading in \texttt{Milo}, config in \texttt{Kura}) rather than in trivial getters/setters or massive monolithic kernels. This confirms that the economic principle of reuse (i.e., copying code that is useful enough to save time but simple enough to adapt) holds true in the IIoT domain as well.

(3) Framework-Driven Cloning (Universal): Similar to how Android frameworks drive cloning in Java \citep{Lopes2017}, our data shows that cloning is heavily caused by domain-specific frameworks (e.g., \texttt{Kura} and \texttt{Kapua} serving as the ``Android of IIoT''). This confirms that ecosystem hubs are the primary vectors for code propagation, regardless of the domain.

\subsection{Implications} \label{Implecation} 
Based on the empirical findings, we derive the following actionable implications for different stakeholders in the Eclipse IIoT software ecosystem.

\textbf{For Developers: Optimizing Reuse Efficiency and Stability.} Individual developers can refine their coding practices to manage technical debt at the source: (1) Implement Real-time Detection for Intra-commit Clones: Our analysis reveals that an average of 31.76\% of clone pairs originate within a single commit, with some projects like \texttt{Milo} reaching as high as 74.53\% for Type-1 clones. Developers should integrate clone-checking tools into pre-commit hooks or local CI pipelines to identify and refactor identical logic before the code enters the shared repository, thus preventing immediate technical debt. (2) Prioritize Type-3 Clone Awareness during Adaptive Reuse: Type-3 clones dominate the IIoT ecosystem, accounting for 56.72\% of total cloned lines. This indicates a pattern of ``adaptive reuse'' where logic is copied and then modified to fit specific hardware or protocol constraints. Developers must maintain explicit documentation or tracking for these inexact clones to ensure that bug fixes in the original logic are correctly propagated to the adapted versions.

\textbf{For Maintainers and Architects: Strategic Governance and Quality Control.} Maintainers and architects can use these insights to optimize long-term maintenance resource allocation: (1) Target Ecosystem Hubs for Rigorous Quality Assurance: Projects designed for data services, specifically \texttt{Kura} and \texttt{Kapua}, exhibit the highest density of cross-project clones in the Java ecosystem. Architects of projects that depend on these hubs (e.g., \texttt{Arrowhead} and \texttt{Milo}) should treat them as critical dependencies, as quality issues or protocol changes in these hubs are likely to propagate through the entire ecosystem via cloned code. (2) Assess Refactoring Necessity based on Co-modification Risks: Reviewers noted that refactoring in safety-critical IIoT systems can be risky. Our data shows that overall co-modification is rare (0.17\%). Therefore, maintainers should avoid blind refactoring of stable, non-co-modified clones (like the loading functions in \texttt{Milo}) and instead focus refactoring efforts on clones that demonstrate a history of synchronized changes, balancing code purity with system stability.

\textbf{For Researchers: Investigating Domain-Specific Patterns.} Our study provides a foundation for the academic community to further explore industrial software evolution: (1) Quantify the Impact of Protocol Evolution on Technical Debt: The case of \texttt{VOLTTRON}, which saw clones drop from 466K to 145 after removing incompatible protocols, highlights how protocol migration drives clone spikes. Future research should use Spearman’s rank correlation tests or other test (as applied in this study, where 6 projects showed significant increases with $\rho$ up to 1.00) to model how architectural shifts and standard updates influence the accumulation and reduction of code clone. (2) Develop Context-Aware Detection Tools for Industrial Code: Given the high incidence of Type-3 clones , existing token-based or exact-match tools may be insufficient for IIoT. Researchers should focus on developing detection algorithms that can handle the ``semantic noise'' introduced by hardware-specific modifications and configuration settings, which are prevalent in the Eclipse IIoT software ecosystem.

\section{Threats to Validity}\label{sec_threats}

There are several threats to the validity of the study results. We discuss these threats according to the guidelines in \citep{RuHo2009}. Please note that internal validity is not discussed since we do not study causal relationships.

\subsection{Construct Validity}
Construct validity refers to the extent to which our measurements accurately reflect the theoretical concepts under study. In this study, we used NiCad \citep{Roy2008Nicad} to detect code clones, a tool widely adopted in previous studies. However, different configurations of NiCad may affect the final results, potentially threatening the construct validity of this study. To minimize this threat, we employed settings that have demonstrated high recall and accuracy in existing studies \citep{Sobrinho2021ASystematicBS, Mo2023ExploringTI, Assi2024UnravelingCC, Svajlenko2014TowardsAB}. 

A significant threat lies in the choice of metrics used to quantify clone prevalence and evolution (e.g., $CLOC$, $POC$, and \textit{Clone Pairs}). As noted in Section \ref{dataItemsCollected}, we primarily rely on \textit{Clone Pairs} and \textit{Lines of Code (LOC)}. However, these metrics may introduce biases; for instance, counting pairs can over-represent code blocks that are cloned many times (clone groups), while LOC-based metrics vary depending on code formatting (e.g., braces placement). To mitigate this, we normalize the source code using NiCad's pretty-printing feature before counting lines. Furthermore, by reporting multiple complementary metrics (both absolute pairs and relative percentages), we provide a multifaceted view of clone distribution, reducing the risk of skewing conclusions based on a single, potentially biased indicator.

%\subsection{External Validity}
%External validity refers to the generalizability of our research findings to other contexts. In this study, an important threat to external validity is that we only examined a limited number of projects within a single OSS ecosystem. To address this limitation, we plan to replicate our study in a broader range of projects in the future work. We also consider collaborating with industrial partners to conduct a more comprehensive maintenance risk assessment of a wider array of IIoT software projects. These efforts will improve the external validity of our research findings.

\subsection{External Validity}
External validity refers to the generalizability of our research findings to other contexts. In this study, two important threats to external validity exist.

First, we only examined a limited number of projects within a single OSS ecosystem. While the IIoT domain is important and growing, the findings may not generalize to other software ecosystems or application domains. To address this limitation, we plan to replicate our study in a broader range of projects in the future work. We also consider collaborating with industrial partners to conduct a more comprehensive maintenance risk assessment of a wider array of IIoT software projects.

Second, a notable threat concerns the representation of programming languages in our sample. Our selected projects are predominantly Java (10 projects), with smaller representations of C (4 projects) and Python (1 project). This unbalanced distribution raises concerns about generalizability across different programming languages. The underrepresentation of Python and other languages in our sample limits our ability to draw conclusions about clone prevalence in these languages. Future studies should aim for a more balanced selection of programming languages to validate whether the observed clone prevalence hold consistently across different language ecosystems and their associated problem domains. These efforts will improve the external validity of our research findings.

\subsection{Reliability}
Reliability refers to the extent to which the results of a study are consistent when repeated by other researchers. Potential threats are related to data collection, which involves the tools used and the data items collected. The code of the analysis module was regularly reviewed by the second and third authors. In addition, thorough testing, including unit tests and validation against known datasets, was performed to ensure the accuracy of the analysis results generated by the module. The data items were collected with cross validation by the first and second authors. As a result, the threats to the reliability of this study have been mitigated.

%In summary, we have addressed construct validity by using validated NiCad configurations and cross-validating the dataset, external validity by acknowledging the limitations of closed-source projects and planning future work, and reliability by adopting consistent measurement practices. While some threats remain, these measures enhance the overall validity and robustness of our study.

\section{Conclusions and Future Work}\label{sec_conclusion}
In recent years, IIoT has emerged as a pivotal research field. The accelerated development of digital infrastructure and IIoT-related technologies, coupled with strong policy support, has led to an expanding application of innovations such as digital twins, cloud services, and embedded systems in the field of IIoT. Code clone is a common code smell in software projects, extensively studied for its potential negative impact on software maintainability \citep{Lozano2008AssessingTE, Hu2021CodeCloneHrmfulness}. However, there has yet to be any research on the prevalence and evolution of code clones in IIoT OSS systems, as well as whether code clones will impact the maintenance of the IIoT OSS ecosystem.

In this paper, we conducted an empirical study to explore the prevalence and evolution of code clones as well as their co-modifications in Eclipse IIoT OSS projects. Through our study on 15 Eclipse IIoT OSS projects, we obtained the following findings. First, 16.30\% of all code lines in these projects are involved in code clones, which is nearly twice the proportion observed in traditional OSS projects. Second, the proportions of both intra-commit and inter-commit code clones highlight the need for comprehensive refactoring strategies that distinguish between different temporal contexts in the development process - immediate intervention during code creation and submission versus systematic refactoring approaches for long-term maintenance and architectural improvements. Third, the code clones predominantly exhibit an upward or stable trend throughout the software version iteration process, which suggests that as projects evolve, code clones may continue to expand. Fourth, co-modified code clones, though relatively rare, impose a notable burden on software maintenance in Eclipse IIoT OSS projects. These clones can lead to inconsistencies and increased maintenance efforts, underscoring the need for better management strategies. Finally, there are a considerable number of cross-project code clones, in which the co-modifications are very few and their impact on the Eclipse IIoT OSS ecosystem is almost negligible.

We believe that our empirical study can provide a fundamental and comprehensive understanding of the impact of code clones on IIoT OSS projects, as well as provide insightful implications to practitioners and researchers. In the next steps, we plan to: (1) investigate the impact of inter-commit and intra-commit clones on software maintainability, and analyze how these two clone patterns affect code quality and refactoring efforts, and (2) seek collaborations with industrial partners to replicate our study and assess the scalability of our findings in real-world scenarios.
%and our dataset and detection code can also provide data and relevant technical support for future research.

\section*{Data availability}
We have shared the link to our replication package in the reference~\citep{github_repo}.

\section*{Acknowledgments}
This work was funded by the National Natural Science Foundation of China under Grant Nos. 62176099 and 92582203.

\printcredits

\bibliographystyle{cas-model2-names}

\bibliography{references}
\balance
\end{sloppypar}
\end{document}